\documentclass[aps,rmp,twocolumn,groupedaddress,showpacs,floatfix]{revtex4}
\usepackage[dvips]{color}
\usepackage[dvips]{epsfig}
\usepackage{amsmath}
\usepackage{amssymb}
\usepackage[american]{babel}
\usepackage{hyperref}
\usepackage{epsfig}
\usepackage{paralist} 
\newcommand{\rem}[1]{}

\begin{document}

\title{Modeling friction: from nano to meso scales}

\author{Andrea Vanossi}
\affiliation{CNR-IOM Democritos National Simulation Center, Via Bonomea 265, 34136 Trieste, Italy}
\affiliation{International School for Advanced Studies (SISSA), Via Bonomea 265, 34136 Trieste, Italy}

\author{Nicola Manini}
\affiliation{Dipartimento di Fisica, Universit\`a degli Studi di Milano, Via Celoria 16, 20133 Milano, Italy}
\affiliation{International School for Advanced Studies (SISSA), Via Bonomea 265, 34136 Trieste, Italy}

\author{Michael Urbakh}
\affiliation{School of Chemistry, Tel Aviv University, 69978 Tel Aviv, Israel}

\author{Stefano Zapperi}
\affiliation{CNR-IENI, Via R. Cozzi 53, 20125 Milano, Italy}
\affiliation{ISI Foundation, Viale S. Severo 65, 10133 Torino, Italy}

\author{Erio Tosatti}
\affiliation{International School for Advanced Studies (SISSA), Via Bonomea 265, 34136 Trieste, Italy}
\affiliation{CNR-IOM Democritos National Simulation Center, Via Bonomea 265, 34136 Trieste, Italy}
\affiliation{International Center for Theoretical Physics (ICTP), Strada Costiera 11, 34104 Trieste, Italy}

\date{\today}
\begin{abstract}
The physics of  sliding friction is gaining impulse from nano
and mesoscale experiments, simulations, and theoretical modeling.
This colloquium reviews some recent developments in modeling and in
atomistic simulation of friction, covering open-ended directions,
unconventional nanofrictional systems, and unsolved problems.
\end{abstract}

\maketitle

\tableofcontents

\section{Introduction}
\label{introduction:sec}

Frictional motion plays a central role in diverse systems and phenomena
that span vast ranges of scales, from the nanometer contacts inherent in
micro- and nanomachines \cite{Urbakh04} and biological molecular motors
\cite{Bormuth09} to the geophysical scales characteristic of earthquakes
\cite{Scholz98}.
%
Due to its enormous practical and technological importance, the problem
has stimulated progress over the centuries.
Historical figures from Leonardo da Vinci onwards have brought friction
into the field of physics, with the formulation of time-honored
phenomenological frictional laws, which have been referred to as the
Coulomb-Amontons laws.
These statements can be briefly summarized as follows:
(i) frictional force is independent of the apparent area of contact;
(ii) frictional force is proportional to the normal load;
(iii) kinetic friction (the force to keep relative motion at constant
speed) does not depend on the sliding velocity and is smaller than static
friction (the force needed to initiate motion between two contacting bodies
at rest).
Also in the light of a mass of empirical data, serious attempts were made
in the first half of the 20th century toward a microscopic understanding of
these laws
\cite{Bowden50}.
Whereas the basic physics underlying sliding friction -- non equilibrium
statistical mechanics of solids, sheared fluids, and moving surfaces -- is
in principle quite exciting, the field as a whole has (even if with notable
exceptions) failed to attract adequate interest by the physicist until the
last few decades.
A lack of microscopic data, and a corresponding lack of theory, have
perhaps contributed to project an unattractive image of sliding friction.

Three quiet revolutions, of broad nature and unrelated to friction, are
radically changing this state of affairs. First, progress in the
general area of complexity provided new tools to tackle non-equilibrium
disordered systems with interacting degrees of freedom.
Second, and crucial, the
developments in nanotechnology
extended the study of friction and permitted its analysis on
well-characterized materials and surfaces at the nano- and microscale.
Notably the invention of scanning tip instruments of the Atomic Force
Microscope (AFM) family
\cite{Binnig86} has opened {\it nanofriction} as a brand new avenue;
the use of the Surface Force Apparatus (SFA) \cite{Israelachvili92} has led
to the systematic studies of confined mesoscopic systems under shear;
while instruments such as the Quartz Crystal Microbalance (QCM)
\cite{Krim88,Krim96} measure the inertial sliding friction of adsorbate
submonolayers.
Thanks to these methods, a mass of fresh data and information on well defined systems, surfaces,
materials, physical conditions has accumulated in the last two decades
\cite{Carpick97}.
The resulting insights into the atomic size contacts themselves in terms of
chemical interactions and of the elementary processes that are involved in
the excitation and dissipation of energy are changing our perspective.
Third, computer simulations have had a strong boost, also allowed by the
fantastic growth of computer power.
The numerical study of frictional models on one hand, and direct atomistic
molecular dynamics simulations on the other hand, are jointly advancing
our theoretical understanding.
Invaluable initial reviews of the progress brought about by these
revolutions in our physical understanding of sliding friction can be found
in the books by \textcite{PerssonBook} and \textcite{Mate08}.

Despite the practical and fundamental importance of friction and the
growing efforts in the field, many key aspects of dynamics of friction are
not yet well understood.
Even for the most studied nanoscale systems, such as AFM sliding on
graphite or NaCl surfaces, a microscopic mechanism of friction is still
lacking, and experimental observations (for instance, velocity and
temperature dependencies of friction) have been rationalized within
simplified models including empirical parameters \cite{Riedo03, Jansen10,
  Barel11}.
Fundamental theory is still difficult in all fields of sliding friction,
including nanofriction, since the sliding motion generally involves sudden
nonlinear stick-slip events, that cannot be treated within traditional
theoretical
approaches such as linear-response theory and hydrodynamics.
Experiments in tribology have long suffered from the inability to directly
observe what takes place at a sliding interface.
Although AFM, SFA and QCM techniques have identified many friction
phenomena on the nanoscale, many interpretative pitfalls still result from
indirect or ex situ characterization of contact surfaces.
In this Colloquium we will briefly cover some aspects, progress and
problems in the current modeling and simulation of sliding friction, from
nano to mesoscale. In nanoscale friction we consider systems that are
small enough to be treated at the atomistic scale, such as in the AFM
experiments.
For larger systems we need a mesoscopic approach that lies in between the
atomistic details and the macroscopic behavior.
In the spirit of a Colloquium we
intend to draw examples from our own experience to illustrate some of the
concepts and points of interest, growth, and doubt in selected forefront areas.

One of the main difficulties in understanding and predicting frictional
response is the
intrinsic
complexity of highly non-equilibrium processes going on in
any tribological contact, which include detachment and reattachment of
multiple microscopic junctions (bonds) between the surfaces in relative
motion while still in contact \cite{Urbakh04, Bormuth09, Gerde01}
%
Therefore friction is intimately related to instabilities that occur on a
local
microscopic scale, inducing an occasional fast motion of the corresponding
degrees of freedom even if the slider's center-of-mass velocity is
extremely small.
Understanding the physical nature of these instabilities is crucial for the
elucidation of the mechanism of friction, as we will emphasize below.

Sliding friction has been addressed following different types of theoretical
approaches: ``minimalistic'' models (MM)
atomistic molecular dynamics (MD) simulations, mesoscopic multicontact models
and phenomenological rate-state (RS) models.

MMs are discussed in Sec. \ref{models:sec}. They provide an intermediate
level description between atomic scale physics and macroscopic
phenomenological approaches like RS models, focusing on a small number of
relevant degrees of freedom which describe the sliding motion, and exhibit
instabilities during stick-slip.
Applications of MMs provided explanations for phenomena of high complexity
[see, e.g., \cite{Muser03,Vanossi07}].
%
On the whole, MMs are playing a major role in rationalizing the wealth of
nano and mesoscale friction data produced over the last decades.

%

Atomistic MD simulations, discussed in Sec. \ref{MD:sec}, have a wide range
of applicability in nanoscale friction, and have reached a high level of
rigor and accuracy \cite{Robbins01}.
But, as discussed in Sec.~\ref{sizeissues:sec}, they are mostly limited to
time and length scales which are too short to emulate many tribological
phenomena.
An important issue, therefore, is how to reduce the large-scale,
many-parameter MD equations to simpler mesoscale descriptions based on
fewer degrees of freedom.

Multicontact models, discussed in Sec.~\ref{multicontact:sec}, provide such
a simplified description in terms of dynamical formation and rupture of
elastically coupled contacts.
At the largest macroscopic scale, phenomenological RS models simplify the
description even further, introducing one or two dynamical equations with
coefficients chosen to fit experimental quantities and then used to
describe a wide range of observed frictional behavior, such as the
transition between stick-slip (regular or chaotic) and smooth sliding
\cite{Carlson96}, and variations of friction for a sudden change of velocity
\cite{Dietrich79, Ruina83}.
RS models are often the best available approaches to describe macroscopic
friction in the ordinary world, from the microsize \cite{Baumberger99} to
larger and larger scales \cite{Scholz98}.
However, most of the ``state variables'' in RS models cannot be easily
related to physical system properties, a fact that limits the insight and
predictive power of these models.
This Colloquium is limited to nano and mesoscale frictional modeling, and
will not further deal with RS models; the latter are well covered for
example by \textcite{Marone98}.
Instead, Sec.~\ref{special:sec} will present theoretical case studies with
model descriptions of a few examples of nanofrictional phenomena, such as
electronic friction, magnetic dissipation, carbon nanotube tribology.
For a closer and fresh perception of where the field stands we will also
include a partial
and incomplete list of problems that, in our perspective, still stand out
for future theoretical study in friction and nanofriction.

\section{Simple nanofrictional models}
\label{models:sec}

\subsection{The Prandl-Tomlinson model}

\begin{figure} \centering
\includegraphics[width=0.48\textwidth,clip=]{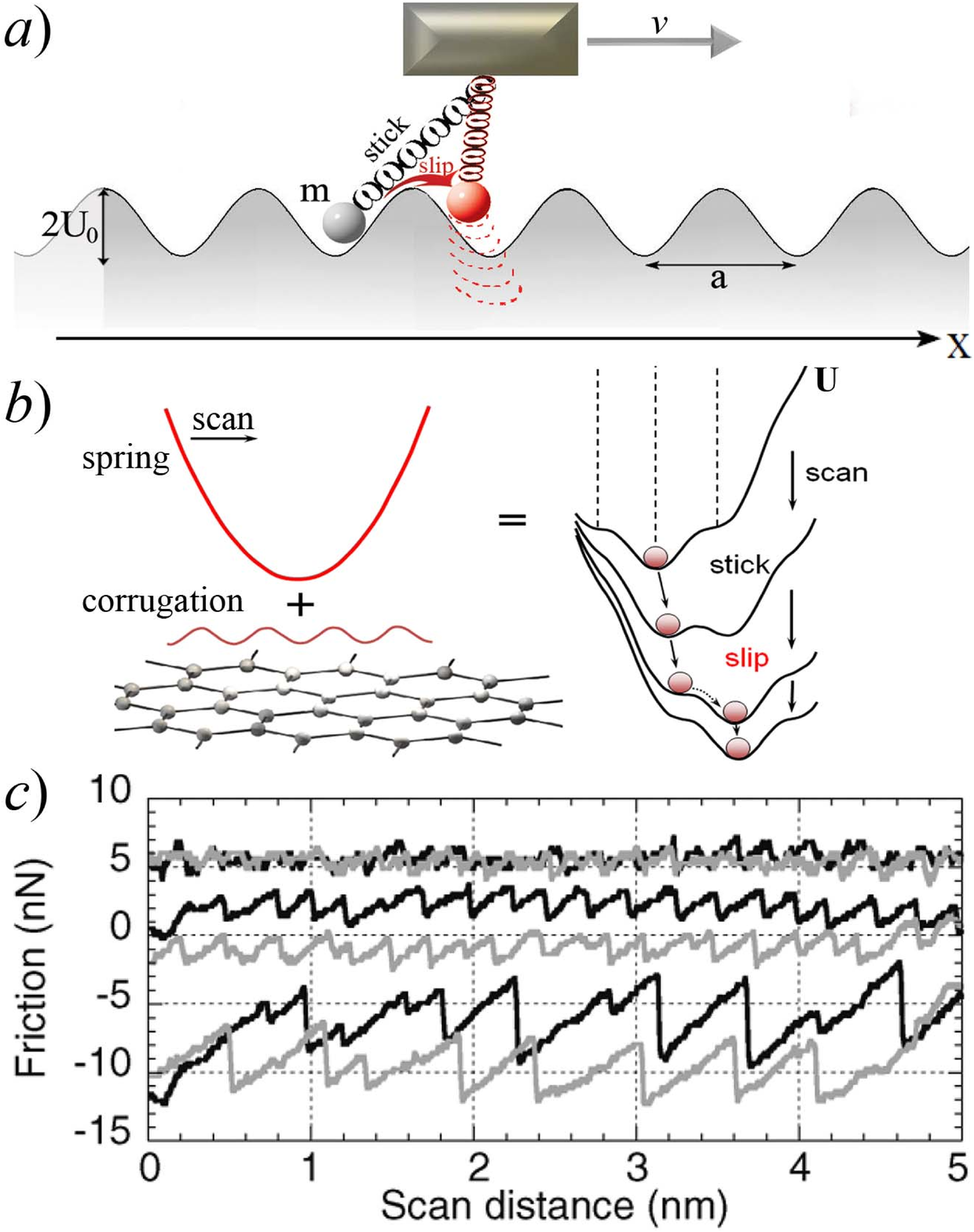}
 \caption{\label{fig:modelPT}
Stick-slip in
(a) a cartoon of the PT model;
(b) Energy landscape for a soft spring (low K). The total potential 
(harmonic spring $+$ sinusoidal substrate) exhibits
different metastable minima, giving rise to the stickslip behaviour;
(c) a representative experimental friction pattern, for increasing
  load. Lateral force vs position traces demonstrate transitions from
  smooth sliding (top) to single (middle) and mostly double slips
  (bottom). From \cite{Medyanik06}.
Similar patterns can be generated within the PT model.
}
\end{figure}

The Prandtl-Tomlinson (PT) model \cite{Prandtl28, Tomlinson29}, which we
discuss here in some detail, is the most
successful and influential MM so far suggested for description of nanoscale
friction.
In particular it addresses friction force microscopy (FFM) where friction
forces are measured by dragging an AFM tip along a surface.
Qualitative conclusions drawn with this model provide guidance to
understand friction at the nanoscale, that often retain their validity in
more advanced models and MD simulations.

PT assumes that a point mass $m$ (mimicking e.g., the AFM tip) is dragged
over a one-dimensional sinusoidal potential representing the interaction
between the tip and a crystalline substrate.
The point tip is pulled by a spring of effective elastic constant $K$,
extending between the tip position $x$ and the position of the microscope
support stage, that is driven with a constant velocity $v$ relative to the
substrate, see Fig.~\ref{fig:modelPT}a.
Thus the total potential experienced by the tip consists of two parts: (i)
the tip-substrate interaction, and (ii) the elastic interaction between the
tip and the support, and can be written as
\begin{equation}\label{tom1}
U\left( x,t \right)
= {U_0}\cos \left( {\frac{{2\pi }}{a}x} \right)
+ \frac{K}{2}{\left( {x - vt} \right)^2}
\end{equation}
where $U_0$ is the amplitude, $a$ is the periodicity of the tip-substrate
potential.
Note that in an AFM experiment the real ``spring constant'' mimicked by $K$
in the PT model is not only due to the torsional stiffness of the cantilever
but includes also the contribution from the lateral stiffness of
the contact.
There is no attempt in the model to describe realistically the
energy dissipation into the substrate (Joule heat) and all dissipation
is described by a viscous-like
force $-m \gamma \dot x$ , where $\gamma$ is a damping coefficient.
The instantaneous lateral friction force measured in FFM experiments reads
$F = - K\left( {x - vt} \right)$, and the kinetic friction $F_k$ is the
time average of $F$.

The PT model predicts two different modes for the tip motion, depending on
the dimensionless parameter $\eta = 4\pi ^2 U_0/(K a^2)$, which
represents the ratio between the stiffnesses of the tip-substrate potential
and the pulling spring.
When $\eta < 1$ the total potential $U(x)$ exhibits only one minimum and
the time-dependent sliding motion is smooth; for $\eta > 1$ two or more
minima appear in $U(x)$, and the sliding is discontinuous, characterized by
stick-slip, Fig.~\ref{fig:modelPT}b. The value $\eta = 1$  represents the transition from smooth sliding
to slips by one lattice site (single-slip regime).

Physically, stick-slip motion corresponds to jumps of the tip between
successive minima of $U(x)$, due to instabilities induced by the driving
spring
($\partial U/\partial x = 0,\;{\partial ^2}U/\partial {x^2} = 0$).
%
%
Close to the inflection point the height of the barrier preventing the tip sliding
decreases with the applied force, as $\Delta E \propto {\left( {const - F} \right)^{3/2}}$  \cite{Dudko02,Sang01,Maloney}
This type of externally induced topological change in a free energy landscape is known as a fold catastrophe,
and it has been found in many driven systems, including superconducting quantum interference devices \cite{Kurkijarvi,Garg},
mechanically deformed glasses \cite{Johnson95} and  stretched proteins \cite{Lacks,Berkovich10}.
The simulation results obtained for diverse systems show that the fold catastrophe scaling is in fact accurate
not only in the immediate vicinity of the inflection point but over reasonably large intervals of loads.

The possibility of slips of higher multiplicity (multiple-slip regime)
occurs for larger values of $\eta > 4.604$  \cite{Medyanik06}.
It should be noted that this is a necessary but not
sufficient condition to observe multiple slips, since the observed dynamics
depends also on the damping coefficient $\gamma$.
In particular, for $\eta > 4.604$ one can distinguish between the overdamped
regime of motion, $\gamma > \sqrt{U_0/m}\; 2\pi/a$, where
the tip jumps between nearest-neighbor minima of the potential, and the
underdamped regime, $\gamma < \sqrt{U_0/m}\; 2\pi/a$,
where the tip may perform multiple slips over a number of lattice sites and
even overshoot the lowest well of the potential $U(x)$.
In that case the minimal spring force reached during stick-slip
oscillations is negative.

The elastic instability occurring for $\eta > 1$ results in a nonzero value
of the low-velocity kinetic friction that is given by the energy drop from
the point of instability to the next minimum of the potential divided by
$a$ \cite{Helman94}.
For $\eta < 1$ this instability does not exist, friction is viscous, and
$F_k \to 0$ for $v\to 0$.
%
The emergence of static friction can be interpreted as the arousal of a
saddle-node bifurcation as a function of $\eta$, realizing a sort of
fold-catastrophe scenario \cite{Gilmore81}.
\rem{

HERE OR NEAR HERE

TASK 2a:

ALLUDE TO  GENERAL INSTABILITY SCENARIO, POSSIBLY WITH PICTURE
NICOLA / ANDREA / ERIO
[Andrea's suggestion: Weiss-Elmer's paper, pag. 3 & 4]

(a) An essential feature of friction (especially at low T, but not
exclusively) is the bi- (or multi-) stability corresponding to an
underlying (possibly complicated) energy landscape. This property is
responsible to the emergence of a finite static friction coefficient
and for many other frictional properties that follow from the
activated nature of the problem. In this context, discussing the
saddle-node bifurcation (fold-catastrophe) scenario and scaling,
including some graphical representation, can greatly help some readers
to appreciate the nature of the problem. This will also make the
appearance of the power 2/3 in the temperature dependence of dynamic
friction in Eq. (5) and the discussion around it far less mysterious.
}

Note that in real systems at finite temperature, hysteresis and dissipation
must always disappear in the zero-speed limit of adiabatic sliding, where
stick-slip instabilities are preempted by thermal fluctuations.
This regime, sometimes termed thermolubricity, is addressed in
Sect.~\ref{thermalPT:sec}, particularly by Eq.~\eqref{slowregime}.

In experiment, the effective value of the PT parameter $\eta$ can be
controlled by the variation of the normal load on the contact, which
changes the potential corrugation $U_0$ more than the contact stiffness.
FFM experiments at low normal loads indeed demonstrated smooth sliding with
ultralow friction, connected to the absence of elastic instabilities
\cite{Socoliuc04,Medyanik06}.
At higher loads instead, ``atomic'' stick-slip took place with the atomic
periodicity of the substrate lattice, while increasing load (corresponding
to increasing $U_0$) further led to a multiple slip regime as predicted by
the PT model, see Fig.~\ref{fig:modelPT}c.

\subsection{Extensions of the Prandtl-Tomlinson model}

Several generalizations of the original, one-dimensional PT model have
marked new steps toward understanding and implementation of frictional
phenomena.
These extensions included considerations of:
\begin{compactitem}
\item
  two-dimensional structure of surfaces that led to the introduction of
  frictional imaging of interfaces \cite{Gyalog95, Prioli03, Fusco04,
    Fusco05};
\item
  thermal fluctuations that allowed to understand an origin of velocity
  dependence of friction and introduced a new regime of friction, named
  ``thermolubricity'' \cite{Gnecco00, Sang01, Dudko02, Riedo03, Reimann04,
    Krylov05};
\item
  coupling between normal and lateral motion of the slider \cite{Rozman98a,
    Zaloj99} that led to a new approach to control friction and wear by
  modulating the normal load \cite{Socoliuc06, Lantz09};
\item
  flexibility of the AFM tip apex that led to a predictions of new regimes
  of motion exhibiting complex stick-slip patterns \cite{Krylov06,
    Tshiprut08}.
\end{compactitem}

Deferring some of these points bearing contact with the Frenkel-Kontorova
model, we focus first on the effect of temperature on friction.

\subsection{Thermal and velocity effects on nanoscale friction}
\label{thermalPT:sec}

The main aspects of thermal effects on friction were considered in the
pioneering work by \textcite{Prandtl28}.
Thermal effects can be incorporated into the model \eqref{tom1} by adding a
thermal random force $\hat f(t)$ to the conservative force between the
slider and substrate and the damping term $-m\gamma \dot x$.
Then the tip motion is described by the following Langevin equation
\begin{equation}\label{tom2}
m\ddot x + m\gamma \dot x =
- \frac{\partial U\left(x,t\right)}{\partial x} + \hat f(t)
\end{equation}
The random force should satisfy the fluctuation-dissipation theorem; as
usual, it is chosen with zero mean $\left\langle {\hat f\left( t \right)}
\right\rangle = 0$ and $\delta$-correlated:
\begin{equation}\label{tom3}
\left\langle {\hat f(t)f(t')} \right\rangle
= 2m\gamma \, k_B T\,\delta(t - t')
\,,
\end{equation}
where $k_B$ denotes the Boltzmann constant and $T$ temperature.
The random forces and the damping term arise from interactions with phonons
and/or other fast excitations that are not treated explicitly.

In the thermal PT model, Eqs.~\eqref{tom2} and \eqref{tom3}, beside the
PT-parameter $\eta $, thermal fluctuations bring out a new dimensionless
parameter $\delta$ representing the ratio between the pulling rate $v/a$
and the characteristic rate of thermally activated jumps over the potential
barriers, $\omega_0 \exp\left( - U_0/k_BT \right)$, where $\omega_0$ is the
attempt frequency \cite{Krylov05}.
As a result one should distinguish between two regimes of motion: ({\it i})
$\delta \ll 1$, regime of very low velocities or high temperatures
(typically $v<1$~nm/s at room temperature), where the tip has enough time
to jump back and forth across the barrier, and ({\it ii}) $\delta \gg 1$,
stick-slip regime of motion, where thermal fluctuations only occasionally
assist the tip to cross the barrier before the elastic instability is
reached.
In these two regimes the following expressions for kinetic friction have
been suggested \cite{Sang01, Dudko02, Krylov05}:
\begin{equation}\label{slowregime}
F_k\left( {v,T} \right) =
\alpha \left( T \right)v + O\left( {{v^3}} \right),
\quad
\delta\ll 1
,
\end{equation}
\begin{eqnarray}\label{fastregime}
F_k\left( {v,T} \right) =
F_0 &-& b T^{2/3} \ln^{2/3}\left( B\frac{T}{v} \right),
\\\nonumber
&&
\delta\gg 1 \ {\rm and}\ v < BT
.
\end{eqnarray}
Here $F_0$ is the athermal ($T=0$) low-velocity limit of friction,
$\alpha(T) \propto (K/\omega _0) \left( U_0/(k_BT) \right)\exp\left(
U_0/(k_BT) \right)$ is the equilibrium damping felt by the tip that is
independent of the ad-hoc damping coefficient $\gamma$, and $b$, $B$ are
positive constants which depend on $m$, $K$, $a$, $U_0$ and $\gamma$ but
not on $v$ and $T$.
Equation \eqref{slowregime}, describing the slow friction regime called
thermolubricity \cite{Krylov05}, corresponds to the linear-response regime,
while Eq.~\eqref{fastregime} has been derived assuming that thermally
activated depinning still occurs in the vicinity of the athermal
instability point.
The velocity and temperature dependencies of friction force predicted by Eq. \eqref{fastregime}
result from the fold catastrophe scaling of the potential barriers, $\Delta E \propto {\left( {const - F} \right)^{3/2}}$,
which has been discussed in the previous section.
Furthermore, in between the two regimes described by Eqs. \eqref{slowregime} and \eqref{fastregime}
one should observe a logarithmic dependence of $F_k$ on velocity.
However, it is very difficult to distinguish between
$\left[ {\ln \left( v \right)} \right]^{2/3}$
and simple ${\ln \left( v \right)}$
behavior in experiments and numerical simulations \cite{Muser11}.
The logarithmic (or $\left[ \ln(v) \right]^{2/3}$) regime tends to span
many decades, until $v$ becomes so large that the inertial or viscous-like
effects set in.
The ${\left[ {\ln \left( v \right)} \right]^{2/3}}$ dependence of the
average rupture force has been also found in single-molecule unbinding
experiments where the energy landscape of complex biomolecules is probed by
applying time-dependent forces \cite{Dudko03}.

The theoretical framework outlined above has explained a number of FFM
experimental results on single crystal surfaces \cite{Gnecco00, Riedo03,
  Stills03}.
Furthermore, the statistical distribution of friction forces was measured
to match predictions from the PT model \cite{Schirmeisen05b}.
These results provide strong evidence that atomic stick-slip in FFM is
attributable to thermally activated slip out of a local minimum as
described by the PT model.
Thermally activated stick-slip friction is only seen in MD at sufficiently
low speeds, which are so far only achievable through accelerated MD
\cite{Li11}.
At higher speeds, friction is mostly determined by dissipative athermal
dynamical processes, which correspond to a fundamentally different regime
of sliding.
This limits severely the regime of validity of comparisons of the PT model
with MD simulations.

Equations~\eqref{slowregime} and \eqref{fastregime} also predicts that
kinetic friction should decrease with increasing temperature \cite{Sang01,
  Dudko02, Steiner09}.
Thermal excitations in fact help overcome energy barriers and reduce the
stick-slip jump magnitude, so that nanofriction should decrease with
temperature provided no other surface or material parameters are altered by
temperature \cite{Szlufarska08}.
Up to now, most FFM measurements have been performed at room temperature,
so that the temperature dependence of nanoscale friction has rarely been
addressed in experimental work.
Recent experimental results \cite{Schirmeisen06, Zhao09, Barel10a,
  Barel10b}, however, strongly disagree with the predictions of
Eqs.~\eqref{slowregime} and \eqref{fastregime}.
Friction forces exhibit a peak at cryogenic temperatures for different
classes of materials, including amorphous, crystalline, and layered
surfaces. Can this effect be explained within the PT model?
Recent analysis of the thermal PT model \cite{Tshiprut09, Fajardo10}
demonstrated that the friction force may indeed exhibit a peak in the
interval of temperatures corresponding to a transition from a multiple-slip
regime of motion, at low $T$, to the single-slip regime at higher $T$.
In this picture, interplay between thermally activated jumps over potential
barriers and the reduction of the slip spatial extension with $T$ may lead
to a nonmonotonic temperature dependence of friction.
However, the PT model fails to reproduce the observed features of the
temperature and velocity dependencies of kinetic friction, and of the force
traces measured with atomic resolution.

\subsection{The Frenkel-Kontorova model}
\label{FK:sec}

The basic model describing the sliding of crystalline interfaces is the
one-dimensional Frenkel-Kontorova (FK) model, see \textcite{Braunbook} and
references therein.
First analytically treated by \textcite{Dehlinger29} and then introduced to
describe dislocations in solids \cite{Frenkel38,Kontorova38a,Kontorova38b},
the FK model found subsequently a wide area of applications, in particular,
in surface physics, where it is often used to unravel the physical behavior
of adsorbed monolayers, specifically to address competing incommensurate
periodicities.

\begin{figure} \centering
\includegraphics[width=0.48\textwidth,clip=]{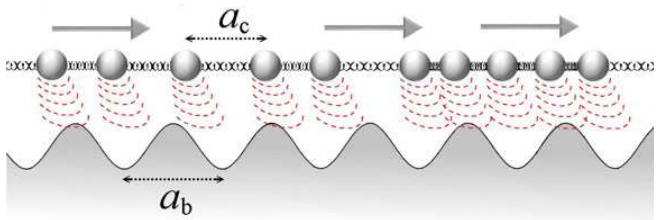}
 \caption{\label{fig:FKmodel}
 A sketch of the FK model with the two competing lengths: interparticle and
 substrate periodicities.
}
\end{figure}

The standard FK model Hamiltonian is
\begin{eqnarray}\label{FKHamil}
H = \sum\limits_i
\left[ \frac{{{p_i}^2}}{{2m}} + \frac{K}{2}(x_{i + 1}  - x_i  - a_c)^2
+ \frac{U_0}{2} \cos \frac{2\pi x_i}{a_b}  \right],
\end{eqnarray}
describing a 1D chain of $N$ harmonically coupled classical ``atoms''
subjected to a sinusoidal potential, see Fig.~\ref{fig:FKmodel}.
The first term in Eq.~\eqref{FKHamil} is the kinetic energy of the chain,
the second one describes the harmonic interaction of the nearest neighbors
in the chain with elastic constant $K$ and equilibrium distance $a_c$, and
the last term is the interaction of the chain particles with the periodic
potential of magnitude $U_0$ and periodicity $a_b$.
Static friction is probed by driving all atoms with an extra adiabatically
increasing force $F$ until sliding initiates.

\begin{figure}
\centering
\includegraphics[width=0.48\textwidth,clip=]{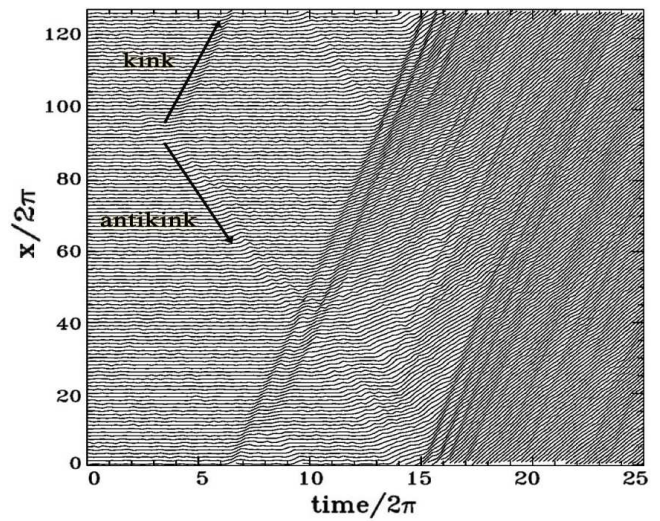}
\caption{\label{kinks}
Detailed behavior (atomic trajectories versus time) at the depinning
transition at a small nonzero temperature of the FK chain with $\theta=1$.
The onset of motion is marked by the creation of one kink-antikink pair.
The kink and antikink move in opposite directions, collide quasielastically
(because of the periodic boundary conditions), and soon a second
kink-antikink pair is created in the tail of the primary kink.
This process repeats with an exponential (avalanche-like) growth of the
kink-antikink concentration, leading to the totally sliding state.
Adapted from \textcite{Braun97}.
}
\end{figure}

Tribological processes in the FK model are ruled by kink (topological
soliton) excitations.
Consider the simplest case of the trivial commensurate ground state when
the number of atoms $N$ coincides with the number of minima of the
substrate potential $M$, so that the dimensionless concentration $\theta
=N/M=a_b/a_c$ is 1.
In this case, adding (or subtracting) one extra atom results
in a chain configuration with a kink (or an antikink) excitation.
After relaxation, the minimum-energy configuration corresponds to a local
compression (or extension in the antikink case) of the chain.
Kinks are important because they move along the chain far more easily than
atoms: the activation energy for kink motion (the Peierls-Nabarro barrier)
is always smaller or much smaller than the amplitude $U_0$ of the substrate
potential.
%
The relevance of topological defects for slip can be understood going back
to the pioneering work by Frenkel on the shear strength of crystalline
solids \cite{Frenkel26}.
Frenkel estimated the ideal plastic yield stress of a crystal as the stress
needed to displace one atomic plane by one lattice spacing.
This calculation leads to an estimate for the yield stress which is of the
order of the shear modulus.
The result is in contradiction with experiments which typically show much
smaller results.
The reason for this discrepancy is rooted in the presence of dislocations
which can be displaced by much smaller stresses, leading to plastic
deformation much earlier than in the perfect-crystal limit.
In ideal conditions, the only barrier to dislocation motion is the
Peierls-Nabarro stress due to the periodic lattice and this is typically
orders of magnitude smaller than the shear modulus.

\rem{
[Andrea's suggestion:
barriera nuda di Peierls-Nabarro per
il kink definita dall'equazione 3.8 nel libro di Oleg,
dove facendo le debite trasformazioni (controllate!)
si dovrebbe avere
E_PN=U_0/2 * 64 * pi^2 * g * exp (-pi^2 sqrt (g))
essendo per noi

By defining a dimensionless interaction strength $g= (a_b / 2\pi)^2 2K /
U_0$, the effective energy barrier $E_{\rm PN}$ for the advancement of a
kink, decreases from $U_0$ at small $g$ to $\simeq U_0 \, 32 \pi^2 g
\exp(-\pi^2\sqrt{g})$ at large $g$ \cite{Braunbook}

b) In the introduction to the FK model, the authors explain that the
model has emerged in the context of dislocations motion is solids.
They also state that the presence of kinks (i.e. dislocations) makes
sliding much easier. However, the connection to dislocations and their
essential role in explaining the practical shear strength of
crystalline solids and interfaces, as compared to the ideal shear
strength, is not really discussed. I think that including such a
discussion will both make things clearer to the non-specialist reader
and will put the friction issues within a broader context. The authors
can follow Frenkel's calculation for the ideal strength, i.e. the
force needed to move the whole chain simultaneously and then to
discuss the role of stress concentration near the kink/dislocation
core to explain why the Peierls-Nabarro stress/threshold is so much
smaller than the ideal threshold.
}


Because the kinks (antikinks) correspond to extra atoms (vacancies), their
motion provides a mechanism for mass transport along the chain and are thus
responsible for mobility, conductivity, and diffusivity.
The higher the concentration of kinks, the higher will be the system
mobility \cite{Vanossi03}.
When the ground state is commensurate (i.e., $\theta =1$), at nonzero
temperature, the first step to initiate motion in the FK model is the
creation of a kink-antikink pair, see Fig.~\ref{kinks}.

When the elastic layer is of finite extension, kinks are usually generated
at one end of the chain and then propagate along the chain until
disappearing at the other free end.
Each run of the kink (antikink) through the chain results in the shift of
the whole chain by one lattice constant $a_b$.
In the case of a finite film confined between two solids, one may similarly
expect that the onset of sliding is initiated by the creation of a local
compression (kink, misfit dislocation) at the boundary of the contact,
while kink's motion is the basic mechanism of sliding.

A crucial role in the FK model is played by incommensurability and the {\it
  Aubry transition} \cite{Peyrard83} connected with it.
Let the substrate period $a_b$ and the natural period of the chain $a_c$
be such that, in the limit of infinite system's length, their ratio
$\theta =a_b/a_c$ is irrational.
Roughly speaking, in this case the FK chain acquires a ``staircase''
deformation, with regions of approximate commensurability separated by
regularly spaced kinks (or antikinks if $\theta<1$).
If there is a nonzero probability to find particles arbitrarily close to the
maximum potential energy $U_0$ these kinks are unpinned and mobile,
otherwise they are pinned \cite{Floria96}.
For a fixed amplitude of the substrate potential $U_0$, the FK ground state
undergoes a transition between these two states (the Aubry transition) at a
critical value $K= K_c$ of the chain stiffness.
$K_c$ depends dramatically and discontinuously on the incommensurability
ratio $a_b/ a_c$ defining the interface.
In particular, it has been proven that $K_c$ takes the minimal possible
value equal to $\approx 1.0291926$ [in units of $2 U_0 (\pi/a_b)^2$]
for the ratio equal to the irrational golden mean $a_b / a_c = (1+
\sqrt{5})/2$ \cite{Braunbook}.
From a physical point of view, this means that for $K>K_c$ there is a
continuum set of ground states that can be reached adiabatically by the
chain through nonrigid displacements of its atoms at no energy cost ({\em
  sliding mode}).
On the other hand for $K<K_c$, the atoms are {\it all} trapped close to the
minima of the substrate potential and thus require a finite energy per kink
(equal to the Peierls-Nabarro barrier) to move over the corrugated
substrate.
Thus, for incommensurate contacts above the Aubry transition ($K>K_c$), the
kinks are mobile, chain sliding is initiated by even the smallest driving
force and, accordingly, the static friction force vanishes, $F_s=0$ -- the
chain sliding is {\em superlubric}.
On the other hand, below $K_c$ the two incommensurate 1D surfaces are
locked together due to pinning of the kinks that separate local regions of
common periodicity, and in this case we expect stick-slip.

The kinetic friction properties of the FK model \cite{Strunz98a,Strunz98b}
are probed by adding a (e.g. Langevin) thermostat as described for the PT
model above.
Even where (above the Aubry transition) $F_s=0$ the kinetic friction force
$F_k$ is nonzero, because the dynamics at any finite speed results in the
excitation of phonons in the chain.
Note also that a {\it finite-size} $T=0$ FK model is always statically
pinned, even for an irrational value of $a_b / a_c$ because of the locking
of the free ends of the chain.
However an Aubry-like transition, exhibiting a symmetry-breaking nature,
can still be defined \cite{Braiman90,Benassi11a,Pruttivarasin11}.
At finite $T$, pinning can be overcome by thermal fluctuations,
which can initiate sliding even in the most-pinned state, the fully
commensurate one, see Fig.~\ref{kinks}.
Finally, we remark that friction in the dynamically driven FK model
describes fairly just the onset of sliding of a crystalline contact
\cite{Hammerberg98}, while it cannot account for the highly inelastic
plastic or quasi-plastic deformations of the surfaces characterizing
real-life friction experiments.

\subsection{Superlubricity}

Superlubricity is the phenomenon in which two incommensurate periodic
surfaces may slide in dry contact with no atomic scale stick-slip
instabilities which, as discussed above, are the main source for energy
dissipation.
Its physical origin is first of all that the energy of two interacting
infinite incommensurate systems is independent of their relative position;
and secondly that if they are hard enough they will slide without
stick-slip.

Vanishing static friction has been first obtained within the FK model in
the pioneering work of \textcite{Peyrard83} for mutually incommensurate
periodicities, and sufficiently hard infinite lattices.
Later, \textcite{Hirano90, Hirano93b, Shinjo93} predicted that for infinite
incommensurate contacts also the kinetic friction should vanish, and called
this effect superlubricity.
In these conditions, the lateral corrugation forces between two
non-matching, rigid crystals cancel out systematically, so that the kinetic
friction of externally driven solid vanishes at zero speed, and is
dramatically reduced even at finite speed.

The term superlubricity has been criticized as misleading, since it might
wrongly suggest zero friction in the sliding state in analogy to
superconductivity and superfluidity.
Instead, incommensurability of periodic interfaces cancels only one of the
channels of energy dissipation, that originating from the low-speed
stick-slip instability.
Other dissipative processes, such as the emission of sound waves, still
persist, and therefore even in the case of complete incommensurability the
net kinetic friction force does not vanish.
Nonetheless, in the superlubric regime one expects a substantial reduction
of the friction force relative to a similar, but commensurate case.

\begin{figure} \centering
\includegraphics[width=0.48\textwidth,clip=]{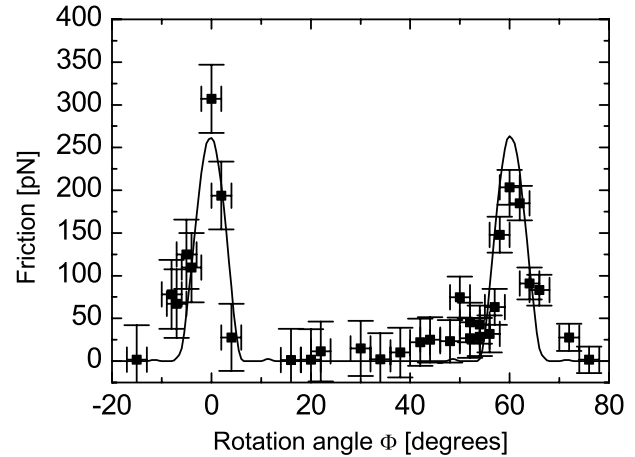}
 \caption{\label{fig:flakesuperlub}
  The data points show the average friction force versus the rotation angle
  measured by \textcite{Dienwiebel04}.
  The curve through the data points shows the calculated friction force
  from a generalized PT model for a symmetric 96-atom flake.
  From \textcite{Verhoeven04}.
}
\end{figure}

\begin{figure} \centering
\includegraphics[width=0.48\textwidth,clip=]{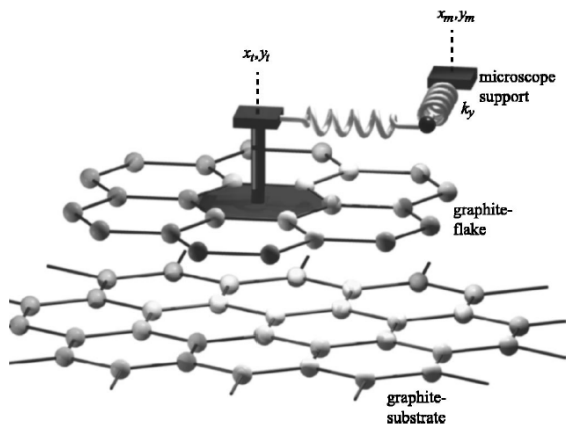}
 \caption{\label{fig:modifiedtom}
  The modified PT model used in the simulations of superlubricity.
  A rigid flake consisting of $N$ atoms (here $N=24$) is connected by an
  $x$ spring and a $y$ spring to the support of the microscope.
  The support is moved in the $x$ direction. The substrate is modelled as
  an infinite rigid single layer of graphite.
  From \textcite{Verhoeven04}.
}
\end{figure}

Detailed experimental studies of superlubricity were recently performed by
\textcite{Dienwiebel04, Verhoeven04, Dienwiebel05}, who measured friction
between a graphite flake attached to the FFM tip and an atomically flat
graphite surface.
Super-low friction forces ($<50$~pN) are found for most relative
orientations of the flake and the substrate, for which the contacting
surfaces find themselves in incommensurate states (see
Fig.~\ref{fig:flakesuperlub}).
For narrow ranges of orientation angles corresponding to commensurate
contacts, stick-slip motion was observed and friction was high (typically
$250$~pN).
A few earlier experiments \cite{Hirano91, Sheehan96} also provided
indications of superlubricity in dry friction.

These observations of superlubricity can be described within a generalized
PT model treating the graphite flake as a rigid finite lattice, composed of
hexagonal carbon rings, as shown in Fig.~\ref{fig:modifiedtom}.
The interaction potential is obtained by summing the pairwise
interactions between carbon atoms in the flake and in the graphite surface.
The resulting flake-surface potential ${U_{\rm flake}}\left(x_c,y_c,\phi
\right)$ depends on the position of the center of mass of the rigid flake
given by the two-dimensional coordinate ${\bf{r}}_c = ({{x_c},{y_c}})$, and
on the orientational (misfit) angle $\phi$ of the flake relative to the
surface lattice.
The motion of the flake attached to the FFM tip and driven along the
surface is described by the PT equation \eqref{tom2} where the sinusoidal
tip-surface potential is replaced by the more complex potential $U_{\rm
  flake}\left(x_c,y_c,\phi \right)$.
Assuming that the flake is rotationally locked (i.e.\ $\phi$ is constant)
the angular dependence of average friction force is in agreement with
observations: the friction exhibits narrow peaks of high friction with
stick-slip motion around the values of $\phi$ corresponding to the
commensurate configurations of the flake and the surface, which are
separated by wide angular intervals with smooth-sliding ultra-low friction
corresponding to incommensurate configurations \cite{Verhoeven04,Merkle07}.
It should be noted that the angular width of the friction maxima, $\Delta
\Phi $, should depend on the flake size, $\tan \left( {\Delta \Phi }
\right) = 1/D$ , where $D$ is the flake diameter, expressed in lattice
spacings.
Accordingly, the width of friction peaks can be used to estimate the flake
diameter.

Superlubricity between incommensurate surfaces provides a desired
low-friction state essential for the function of small-scale machines.
However, some experiments show that flake superlubricity has a finite
lifetime: it disappears due to a reorientation of the flake into the
commensurate state \cite{Filippov08} as observed in a generalization of
the PT model \cite{Filippov08, deWijn10} and in tight-binding atomistic
simulation \cite{Bonelli09}.

Studies of superlubricity may have important implications for understanding
the macroscopic properties of graphite and other solid lamellar lubricants
which are common solid lubricants \cite{Heimberg01,Rapoport97,Singer98}.
Mainly used as flaky powder, they are applied where liquid lubricants
cannot be used, and show remarkable nanotribological properties which are
still not understood.
Recent MD simulations \cite{deWijn11} demonstrated that two surfaces
lubricated by mobile, rotating graphene flakes may exhibit stable
superlubric sliding as for ideally incommensurate contacts and for surfaces
covered by randomly oriented pinned graphene patches.
%
%
Under humid conditions, the multi-domain surface structures can form
spontaneously due to the capillary forces which fix randomly oriented
flakes at the sliding surfaces, while in vacuum graphite patches are free
to reorient to a high-friction and high-wear regime.
%
This may provide an answer to the long-standing problem of why graphite is
such a bad lubricant in vacuum, and needs the humidity of air to perform
well \cite{Savage48}.

\subsection{Extensions of the Frenkel-Kontorova model}
\label{extensionsFK:sec}

Many relevant generalizations of the FK model have been proposed so far to
cover a large class of frictional relevant phenomena; they mainly consist
of modifications of model interactions or of dimensionality.
For realistic physical systems (as, e.g., atoms adsorbed on a crystal
surface), anharmonicity can be introduced in the chain interatomic
potential, see \cite{Braunbook}.
The main novelties here include effects such as a broken kink-antikink
symmetry, new types of dynamical solitons (supersonic waves), a breakup of
the antikink soliton followed by a chain rupture, a changed kink-kink
interaction.
Likewise, nonsinusoidal periodic substrates, characterized, e.g., by sharp
bottoms and flat barriers \cite{PeyrardRemoissenet}, have been investigated
to address atoms adsorbed on simple metal surfaces.
Complex unit cell substrates \cite{Remoissenet84, Vanossi03}, as well as
quasiperiodic \cite{Vanossi00, vanErp99} and disordered corrugated profiles
\cite{Cule96, Guerra07} have also been considered in simulations.
These deviations from the standard FK potential may lead to qualitatively
different excitations such as different types of kinks, phonon branches,
changes in kink-antikink collisions.
From a tribological point of view, different types of sliding behavior are
to be expected at low-driving forces, when the dynamics is mainly governed
by the motion of kink-like structures.

An important and more realistic generalization of the standard FK chain
with relevant consequences for the resulting tribological properties
(critical exponents, scaling of friction force with system size, mechanisms
of depinning, etc.) involves increasing the dimensionality of the model.
Especially the FK 2D generalized versions of the model \cite{PerssonBook,
Braunbook} are naturally applicable to the description of a contact of
two solid surfaces (i.e., the case of ``dry'' friction), in particular as
is realized in QCM experiments, where 2D monoatomic islands of adsorbate
atoms slide over a periodic crystalline substrate \cite{Krim88}.
These approaches are especially powerful in the investigation of the
transient behavior at the onset (or stopping) of sliding, which is quite
difficult to study in fully realistic 3D models, see e.g.\ \cite{Braun01b}.

Noncontact AFM tips oscillating on top of kink-like adsorbate regions
\cite{Maier08} dissipate significantly more than near in-registry regions.
This mechanism is explained by the higher softness and mobility of
solitonic regions \cite{Gauthier00, Bennewitz00, Loppacher00, Hoffmann01},
and it has been demonstrated by the dynamics of an incommensurate FK chain,
forced and probed by a locally-acting oscillation \cite{Negri10}.

In investigating confined systems under shear, FK-like models with just one
particle \cite{Rozman96a, Rozman96b, Muser02} or an interacting atomic
chain \cite{Rozman97, Rozman98b, Braun05} embedded between two competing
substrates have led to uncover peculiar tribological phenomena related to
stick-slip dynamics or to the appearance of ``quantized'' sliding regimes
of motion \cite{Santoro06, Vanossi06, Vanossi07PRL, Manini07extended,
  Castelli09}.
While some of these phenomena, such as chaotic and inverted stick-slip
motion, two types of smooth sliding and transitions between them, have been
already observed \cite{Drummond01,Drummond03}, others are still waiting for
experimental confirmation.

Last but not least, the combined Frenkel-Kontorova-Tomlinson (FKT) model
\cite{Weiss96,Weiss97} has been introduced including harmonic coupling of
the interacting chain atoms to a sliding body.
This approach resembles the Burridge-Knopoff model, to be detailed in
Sec.~\ref{Larkin:sec}, where, however, an on-site interaction with the
lower body replaces a phenomenological dry friction law (usually a
velocity-weakening one).
The FKT model introduces more degrees of freedom than the PT model, and it
has been used to describe effects of finite size and stiffness of the AFM
tip and of normal load on friction \cite{Igarashi08,Kim09}.
The latter effect has been modeled assuming a linear dependence of the
amplitude $U_0$ of potential corrugation on the applied normal force.
The validity of the FKT model has been tested by 3D MD simulations
\cite{Kim09}, which confirmed the outcome of the model for most of
investigated regimes except the limit of very low stiffness and high normal
load.
Unlike the FKT model in which the breakdown of superlubricity coincides
with the emergence of the metastable states, in the 3D model some
metastable states appear to reduce frictional force leading to nonmonotonic
dependence of force on normal load and tip compliance.

Increasing dimensionality and adding realistic features to the FK model
brings its extensions into closer and closer contact to full-fledged MD
simulations.

\section{Molecular-Dynamics Simulations}
\label{MD:sec}

The simple low-dimensional MMs discussed in Sect.~\ref{models:sec} are
useful for a qualitative understanding of many physical aspects of
friction.
To address subtler features, such as the temperature dependence of the
static friction of a specific interface or the Joule-heat dissipation, one
should go beyond MMs including atomistic structural details of the
interface.
Such an approach is provided by MD simulations.

\begin{figure}
\centerline{
\includegraphics[width=0.48\textwidth,clip=]{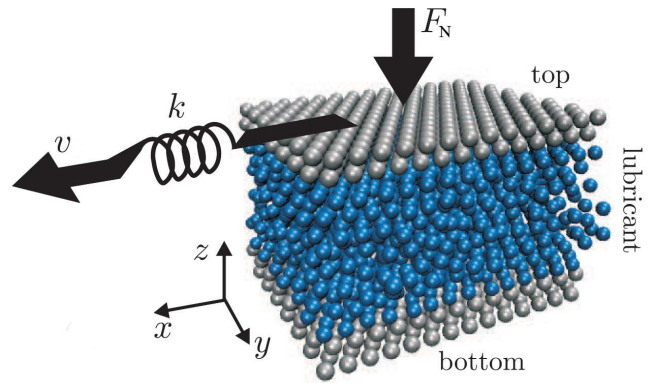}
}
\caption{\label{confined}(color online).
Sketch of a typical MD simulation of a boundary-lubricated interface under
shear.
Periodic boundary conditions are applied in the $x-y$ directions.
}
\end{figure}

\begin{figure}
\centerline{
\includegraphics[width=0.48\textwidth,clip=]{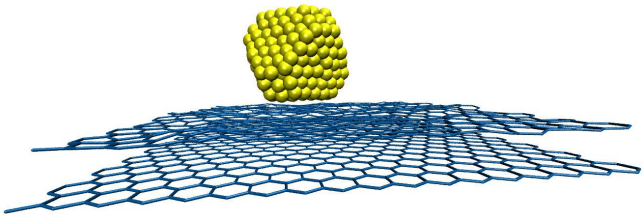}
}
\caption{\label{cluster}(color online).
 A simulated truncated-octahedron Au$_{459}$ cluster sliding with one of
 its $(111)$ facets over a mobile graphite substrate.
 From \textcite{Guerra10}.
}
\end{figure}

Advances in computing hardware and methodology have dramatically increased
our ability to simulate frictional processes and gather detailed
microscopic information for realistic tribological systems.
MD simulations are used extensively in sliding nanofriction to provide
unique insight into the relevant processes, sometimes overturning
conventional wisdom.
They represent controlled computational experiments where the dynamics of
all atoms is obtained by solving numerically Newton (or Langevin) equations
of motion based on suitable interparticle interaction potentials and the
corresponding interatomic forces.
The geometry of the sliding interface and the boundary conditions (e.g.\ as
sketched in Figs.~\ref{confined} and \ref{cluster}) can be chosen to
explore friction, adhesion, and wear.
A thermostat, or other form of damping, is introduced in order to eliminate
the Joule heat to obtain a frictional steady state.
Finally, after specifying the initial coordinates and velocities of the
particles, the classical differential equations of motion are integrated
numerically.

A worthwhile guide to atomistic MD simulations of frictional processes
focusing on fundamental technical aspects (realistic construction of the
interface, appropriate ways to impose load, shear, and the control of
temperature) can be found in the review articles by \textcite{Robbins01}
and by \textcite{Muser06}.
For the general classical MD approach, we refer the reader to the textbooks
by \textcite{Allen91} and by \textcite{Frenkel-Smit96}.

By following the Newtonian dynamics of a system executing sliding for a
significant amount of time, quantities of physical interest such as
instantaneous and average friction force, mean (centre-of-mass) slider
velocity, heat flow, and correlation functions are numerically evaluated.
Unlike standard equilibrium MD simulations of bulk systems, frictional
modeling inherently involves non-equilibrium conditions and a non-linear
dissipative response to the external driving.
A standard practical assumption is to add Langevin terms to Newton's
equations, like in Eqs.~\eqref{tom2}, \eqref{tom3} for the PT model at
finite temperature.
We will return later to this point.

The choice of the appropriate interaction forces between atoms represents
a major problem.
If $U\{R_1, R_2, ...R_N\}$ is the total energy of the system (here say, the
slider plus the substrate), as a parametric function of all atomic
coordinates $\{R_i\}$, the force on atom $i$ is $F_i = - \nabla_{R_i} U$,
perfectly determined once $U$ is known accurately.
Unfortunately this is generally not the case, for $U$ is determined by the
quantum mechanics of electrons -- a much bigger and unsavory problem to
solve.
{\it Ab-initio} MD, e.g.\ of the Car-Parrinello type \cite{Car85}, has not
really been of use so far in sliding friction, mainly because it can handle
only rather small systems, typically hundreds of atoms, for relatively
short times, typically $\ll 1$~ns.
Most MD frictional simulations are therefore based on reasonable empirical
interatomic forces (``force fields''), ranging from relatively
sophisticated energy surfaces accounting for electrons at the
density-functional level or at the tight-binding level \cite{Xu92}, to
angle-dependent many-particle potentials, to simple pairwise potentials
(e.g.\ Lennard-Jones), to basic simple models of elastic springs,
extensions of FK-type formulations.
In practice, several reliable force fields, parameterized to fit different
ranges of experimental data and material combinations, are available in the
literature \cite{Garrison95, Brenner98, amber, Los05, Ghiringhelli05}.
While this allows qualitative atomistic simulations of sliding friction, it
is often far from quantitative.
For example, during such a violent frictional process as wear, atoms may
change substantially their coordination, their chemistry, sometimes their
charge.
Once a specific system is understood after the elaborate development of
satisfactory potentials, the mere change of a single atomic species may
require a painful reparameterization of the interatomic forces.
As a result, systematic frictional studies are a {\it tour de force} if no
suitable set of consistent potentials is already available.
A promising approach consists in the use of the so-called reactive
potentials \cite{Stuart00, Brenner01, Adri01}, capable of describing
chemical reactions and interface wear, with the advantage, for large-scale
atomic simulations, of a good computational efficiency compared to
first-principle and semi-empirical approaches.

\subsection{Thermostats and Joule heat}
\label{thermostat:sect}

In a tribology experiment, mechanical energy is converted to Joule heat
which is carried away by phonons (and electrons in metals).
In a small-size simulation, the excitations generated at the sliding
interface propagate and crowd-up into an excessively small region of
``bulk'' substrate, where they are back reflected by the cell boundaries,
rather than properly dispersed away.
To avoid overheating and in order to attain a frictional steady state, the
Joule heat must therefore be steadily removed.
If this removal is done by means of standard equilibrium thermostats such
as velocity rescaling or Nos\'e-Hoover or even Langevin dynamics, an
unphysical dissipation is distributed throughout the simulation cell, so
that simulated atoms do not follow their real conservative motion, but
rather execute an unrealistic damped dynamics which turns out to affect the
overall tribological properties \cite{Tomassone97}.
Similarly in the PT and FK models, the damping parameter $\gamma$ is known
to modify kinetic and frictional properties, but there is no
clear way to chose the value of $\gamma$.
To a lesser or larger extent, this lamentable state of affairs is common to
all MD frictional simulations.

\begin{figure}
\centerline{
\includegraphics[width=0.38\textwidth,clip=]{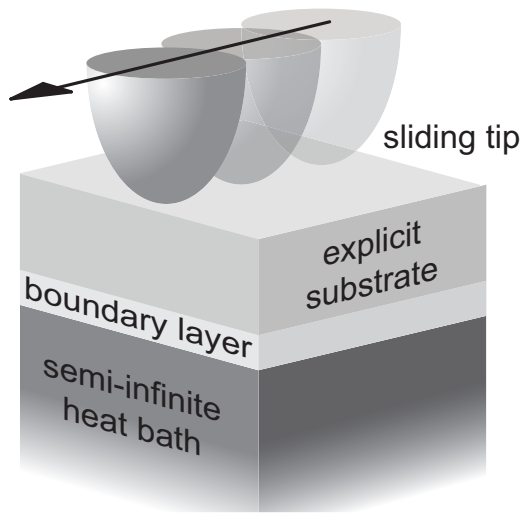}
}
\caption{\label{fig:thermostat}(color online).
Sketch of a MD simulation of friction.
To account properly for heat dissipation, the infinitely-thick substrate is
divided into three regions:
(i) a ``live'' slab comprising layers whose atomic motion is fully
simulated by Newton's equations;
(ii) a dissipative boundary zone, coincident with the deepmost simulated
layer, whose dynamics includes effective damping (e.g., non-Markovian
Langevin-type) terms;
%
(iii) the remaining semi-infinite solid, acting as a heat bath, whose
degrees of freedom are integrated out.
%
%
}
\end{figure}

To solve this problem, one should attempt to modify the equations of motion
inside a relatively small simulation cell so that they reproduce the
frictional dynamics of a much larger system, once the remaining variables
are integrated out.
A recent implementation of a non-conservative
dissipation scheme, based on early formulations by
\textcite{Magalinski59},
and subsequent derivations by
\textcite{Li07,Kantorovich08a,Kantorovich08b}, has demonstrated the correct
disposal of friction-generated phonons, even in the relatively violent
stick-slip regime \cite{Benassi10,Benassi12}.
All atoms near the sliding interface follow plain vanilla Newton's
equation, while the atoms in the deepmost simulated layer, representing the
boundary layer in contact with the semi-infinite heat bath (see
Fig.~\ref{fig:thermostat}), acquire additional non-conservative (and
non-Markovian) terms which account for the time history of this layer
through a memory kernel \cite{Li07,Kantorovich08a}.
Nanofriction simulations exploiting this dissipative scheme have recently
been implemented that conceptually and practically improve over a
traditional Langevin simulation.
Improvement is achieved in particular by adjusting the damping $\gamma$
applied to the simulation boundary layer so as to variationally minimize
the energy back-reflected by that boundary \cite{Benassi12}.

\subsection{Size- and time-scale issues}
\label{sizeissues:sec}

Modern CPUs perform of the order of $10^9$ floating-point operations per
second (FLOPS) per core.
Classical MD can take advantage of medium-scale parallelization, with
fairly linear scaling to approximately $10^2$ cores, thus affording about
$10^{11}$ FLOPS.
As the calculation of the force acting on each atom (usually the dominating
step in a MD calculation) requires, depending on the complexity and range
of the force field, about 10 to $10^2$ FLOPS, the product of the number of
simulated particles $N$ times the number of time-integration steps $N_{\rm
  step}$ per runtime second on a modern medium-size parallel computer is
approximately $N\, N_{\rm step} \simeq 10^{10}$.
With a typical time-step in the fs range, a medium-size simulation
involving $N=10^5$ atoms can progress at a rate of $10^5$~fs per second,
i.e.\ approximately $10^9~{\rm fs} = 1~\mu{\rm s}$ in a simulation day.
This total time scales down for larger systems sizes.

These estimates should be compared with the typical times, sizes, and
speeds of tribology experiments.
If we wish to address macroscopic sliding experiments, typical speeds would
be in the $0.1$ to $10$~m/s range: in $1~\mu{\rm s}$ the slider advances by
$0.1$ to $10~\mu{\rm m}$, i.e.\ approximately $10^3$ to $10^4$ typical
lattice spacings, enough for a fair statistics of atomic-scale events (but
hardly sufficient to gather significant data about phenomena such as the
diffusion of additives or of wear particles within the lubricant, or step-
or defect-related phenomena).
In a nanoscale FFM experiment, however, the tip advances at a far smaller
average speed (i.e. $\simeq 1~\mu{\rm m/s}$) and we can simulate a
miserable $\simeq 1$~pm advancement in a typical run, far too short to
observe even a single atomic-scale event, let alone reaching a steady
state.
Therefore, whenever long-distance correlations and/or slow diffusive
phenomena and/or long equilibration times are expected, MMs will perform
better than fully atomistic MD simulations.
There is nevertheless so much physical insight to be extracted from MD
simulations that it makes sense to run them even at larger speeds than in
AFM of SFA experiments; and indeed, the sliding speed adopted in most
current atomistic MD frictional simulations is in the $0.1$ to $10$~m/s
range.

One of the challenging problems for MD simulations is to account for the
transition from stick-slip to steady sliding.
In SFA and AFM experiments, stick-slip with its associated hysteresis and
large friction generally disappears for speeds larger than $\sim 1 \mu$m/s,
while in MD simulations the transition takes place in the m/s range.
This major discrepancy (up to $\sim 6$ orders of magnitude in speed!)
between simulations and experiments, has been discussed e.g.\ by
\textcite{Braun02a, Luan04, Braun06}, and relates to the effective
spring-force constants and mass distributions, that are hugely different in
the two cases, and much oversimplified in simulations.
Several attempts to fill these gaps rely on methods, including
hyperdynamics, parallel-replica dynamics, temperature-accelerated dynamics,
and on-the-fly kinetic Monte Carlo devised in recent years
\cite{Voter02,Mishin07,Kim11}.
%

Another important aspect present in experiments and largely missed by MD
simulations is the ageing of contacts due to the substrate relaxation.
Ageing can decrease substantially the critical velocity for the transition
from stick-slip to steady sliding.
Contact ageing is also believed to be responsible for the increase of the
static friction force as a function of the contact time.
Direct imaging of contact regions in samples under a normal load show a
logarithmic growth with time \cite{Dieterich94}, leading therefore to an
increasing static friction.
At the phenomenological level, frictional ageing is well described by rate
and state friction laws, widely used in geophysics \cite{Ruina83}, but its
microscopic origin is still debated.
The main mechanisms that have been invoked in the past to explain it are
plastic creep \cite{Heslot94} or chemical strengthening at the interface
\cite{Li11b}.
In a recent paper, AFM was used to explore ageing in nanoscale contact
interfaces, finding supporting evidence for the second mechanism, since
when the contact surface was passivated it showed no more ageing
\cite{Li11b}.
It is however likely that at larger scales and loads plastic creep would
also play an important role.
Beyond its direct relevance for friction, the intriguing issue of contact
ageing occurs in other non-equilibrium disordered systems such as granular
media or glasses.

\subsection{Multiscale models}

Since it is currently impossible to treat atomistically all the
characteristic length scales that mark the dynamical processes entering the
friction coefficient of engineering materials, a rising effort is nowadays
devoted to develop multiscale approaches.
The basic consideration is that unless conditions are very special, all
processes far away from the sliding interface can be described at least
approximately by continuum mechanics, and simulated using finite elements,
allowing for a macroscopic description of elastic and plastic deformation.
The advantage of these continuum-theory methods is that it is possible to
increasingly coarse-grain the system as one moves away from the sliding
contact, thereby highly reducing the computational effort.
Several groups \cite{McGee07,Luan06}
%
combine the atomistic treatment of the interfacial mating region, where
displacements occur on an atomic or larger length scale, with a
coarse-grained or finite-element continuum description elsewhere, where
strains are small and continuous.
The main difficulty lies in the appropriate matching conditions between the
atomistic and continuum regions \cite{E09}.
Since the atomic detail of lattice vibrations (the phonons), which are an
intrinsic part of the atomistic model, cannot be fully represented at the
continuum level, conditions must be met, for example that at least the
acoustic phonons should not be excessively reflected at the
atomistic-continuum interface.
In other words matching at this interface must be such that long wavelength
deformations should transmit with reasonable accuracy in both directions,
which is vital to a proper account of Joule-heat disposal into the bulk.

\subsection{Selected applications of MD in nanotribology}

We survey here some recent 
results from the growing simulation literature, mostly from our groups, and
certainly not providing an adequate 
review of the field.

\subsubsection{Boundary lubrication in confined systems}

Hydrodynamics and elasto-hydrodynamics have been very successful in
describing lubrication by microscopically thin films.
With two sliding surfaces well separated by a hydrodynamically fluid film,
friction is mainly determined by the lubricant viscosity.
The friction coefficient can be calculated using the Navier-Stokes
equations, showing a monotonic increase with the relative sliding velocity
between the two surfaces \cite{Szeri01}.
For small driving velocity and/or high load, the lubricant cannot usually
keep the surfaces apart and solid-solid contact eventually ensues.
Even before full squeezeout, a liquid confined within a nanometer-scale gap
ceases to behave as a structureless fluid.
Pioneering studies of confined systems under shear reveal a sequence of
drastic changes in the static and dynamic properties of fluid films in this
``boundary-lubrication'' regime.

\begin{figure}
\centerline{
\includegraphics[width=0.48\textwidth,clip=]{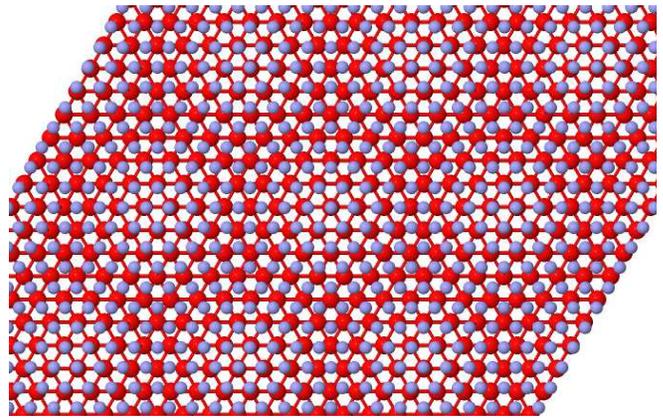}
}
\caption{\label{solid_lub}(color online).
A snapshot of a MD simulation of the 2D solitonic pattern in the boundary
layer of a solid lubricant (clear grey) in contact with a perfect
crystalline surface (dark/red), induced by a 16\% lattice-constant
mismatch.
The Lennard-Jones interaction of this simulation favors in-registry hollow
sites, while unstable top sites mark solitonic regions.
Other layers were omitted for clarity.
%
}
\end{figure}

SFA experiments \cite{Yosbizawa93} and MD simulations \cite{Gao97a,Gao97b}
have both shown clear upward frictional jumps, in correspondence to
squeezout transitions from $N$ to $N$-1 lubricant layers.
The lubricant squeezout for increasing load becomes harder and harder,
corresponding to a (near) crystallization of the initially fluid lubricant
\cite{Persson94a, Persson04a, Tartaglino06}.
%
Owing to layer-by-layer squeezout, there will be frictional jumps for
increasing load.
Friction would not necessarily always jump upward,
because restructuring of the solidified trapped lubricant film, and/or a
switching of the maximum shear gradient from within the lubricant layer
(possibly accompanied by local melting), to the lubricant-slider interface
may take place.
It could jump downward in particular, if lattice mismatch between the
compressed boundary lubricant layer and the rigid substrates jumped, say,
from commensurate to incommensurate, the latter superlubric with a mobile
soliton pattern, as sketched in Fig.~\ref{solid_lub}.
Possibilities of this kind are presently the object of investigation.

Simulations also show that the presence of impurities or defects between
two incommensurate stiff sliding surfaces can, even in a relatively small
concentration, lead to pinning and nonzero static friction \cite{Muser00}.
Defects can destroy superlubricity by introducing mechanical instabilities,
which may occur depending on the system dimensionality, the structure and
relative orientation of the confining walls, and the detail of the
lubricant-wall interactions.

MD investigations of a melting-freezing mechanism in the stick-slip
phenomenology of boundary-lubricated films were carried out by
\textcite{Stevens93,Thompson90}.
Various realistic models for lubrication layers in very specific contexts
have been investigated with extensive MD simulations \cite{Lorenz10a,
  Lorenz10b, Chandross04,Braun06,Chandross08}.

\subsubsection{Sliding of absorbed monolayers}
\label{depinning:sec}

An ideal experimental setup to investigate the molecular origin of
friction is provided by the QCM, where atomically thin systems,
usually just one monolayer
or less of rare-gas 
is
deposited on a 
substrate
resting on a quartz crystal.
When
the crystal surface 
oscillates
strongly enough
so as to dislodge the adsorbate islands
under their own inertial force, 
the sliding
friction 
is revealed by mechanical damping.
Depending on the substrates and on the system interactions at play, the
equilibrium properties of these systems may exhibit distinct structural
geometries, of interest in the field of 2D phase transitions; different
dynamical phases may show up for the driven over-layer, such as pinned
solid, sliding solid, and liquid phases \cite{Krim2007, Krim1991,
  Bruschi2002}.

These experimentally well-characterized systems have also been studied
theoretically and numerically by molecular dynamics simulations
\cite{Cieplak1994, Persson96b, Braun01b, Granato04, Sokoloff1998}.
When the interface is commensurate the static friction is too large to
allow for a massive slip \cite{Cieplak1994}; yet slip is observed
experimentally for a Xe monolayer on a Cu(111) substrate, a system that
forms a commensurate interface \cite{coffey2005}.
A possible explanation is a slip which does not occur coherently but by a
nucleation process in which a bubble slips forward creating a new
commensurate domain as revealed by MD simulations \cite{reguzzoni2010}.
It is possible to estimate the critical radius of this domain and the
energy barrier associated with the nucleation process by the conventional
theory of nucleation, estimating the value of the relevant parameters from
a solution of an effective FK model \cite{Braunbook}.
Thanks to thermally activated nucleation, with the additional help of
impurities and disorder acting as seeds, the monolayer can slip under
lateral forces that are much smaller than those expected for a rigid layer
pinned by the commensurate interface, providing an explanation for the
experimental results.

When the adsorbate is incommensurate and hard, solitons exist already in
the ground state, and their free motion should in principle permit
superlubric sliding.
If however the adsorbate forms islands, as
should  generally
be
the case, perfect
superlubricity is
still
broken by the island's edges, which present a barrier to
the entering and exiting of solitons, necessary for the island to slide.
This island edge pinning is currently under study by N. Varini and
collaborators.

\subsubsection{Extreme temperature/speed conditions}

An advantage of MD is that it can address extreme or otherwise unusual
frictional situations, still unexplored because they may be experimentally
difficult to realize.
One such example are the high ``flash temperature'' regimes caused by local
Joule heating due to wear and other machining or braking conditions
\cite{Bowden50}.
Even in equilibrium but at temperatures close to the melting point, the
outermost layers of a solid substrate generally undergo ``surface melting''
\cite{Tartaglino05}.
In these conditions, AFM nanofriction cannot generally be experimentally
accessed, because the nearly liquefied surface layers jump to contact and
wet the tip long before it reaches nominal contact \cite{Kuipers93}.
However, some solid surfaces, such as NaCl(100), do not melt
\cite{Zykova05}, thus making for an interesting, albeit purely theoretical
so far, case study.
MD simulations predict that high-temperature nanofriction over such a
nonmelting surface would behave very differently depending whether the
tip-surface contact is ``hard'' or ``gentle''.
In the first case, the tip plows the substrate with wear.
The friction coefficient, very large
at low temperature, drops close to the melting
point, when the tip itself provokes local melting, and moves accompanied by
its own tiny liquid droplet,
precisely as in ice skating.
For gentle, low-load, wear free sliding on a hard surface the opposite is
predicted.
Here friction, initially very small, is expected to increase as temperature
is raised close to the melting point where the surface, still solid,
becomes softer and softer due to increasing anharmonicity,
with an analogy to flux lines in type II superconductors. \cite{Zykova07}.

A second example is high-speed nanofriction, as could be expected by a tip
or a surface-deposited nanocluster (Fig.~\ref{cluster}) sliding at large
speed over a smooth crystal surface.
A speed in excess of 1~m/s is many orders of magnitude higher than that of
ordinary AFMs or other nanofrictional systems, and is as yet unexplored.
MD simulations, carried out for the test case of gold clusters sliding over
graphite surfaces, show, besides a standard low-speed
drift sliding
regime, the emergence
of a novel ``ballistic'' sliding regime, typically above 10~m/s
\cite{Guerra10}.
The temperature dependence of the cluster slip distance and time, measuring
its sliding friction, is predicted
to be opposite in these two regimes,
high-speed ballistic sliding and low-speed drift sliding.
The interplay of rotations and translations is crucial to both regimes.
Simulations show that the two are correlated in slow drift but
anti-correlated in fast ballistic sliding.
Despite the deep difference with drift, the speed dependence of ballistic
friction is, like drift, viscous, a useful result whose validity was not
discounted in principle, and which it would be interesting to pursue and
test experimentally.

\subsubsection{Nanomanipulation: pinning vs diffusion}

AFM manipulation of surface-deposited clusters can serve as a useful method
to measure the interfacial friction of structurally well-defined contacts
of arbitrary size and material combinations.
Here, MD simulations are extremely useful in understanding depinning,
diffusion, and frictional mechanisms of clusters on surfaces.
Indeed, one of the remarkable experimental observations of the last decade
concerns the 
unexpected ability of relatively large metal clusters to execute
friction-free motions and even long jumps with size and shape conservation
\cite{Bardotti96,Brndiar11,Dietzel08,Paolicelli08,Paolicelli09}.
%
Gold clusters, comprising typically hundreds of atoms, have been repeatedly
observed to diffuse on highly oriented pyrolytic graphite (HOPG) surfaces
with surprisingly large thermally activated diffusion coefficients already
at room temperature; a similar behavior was reported also for larger
antimony clusters.
MD simulations of the diffusive regime have shown the possible coexistence
of sticking periods, and of long jumps, reminiscent of so-called Levy
flights \cite{Luedtke99, Lewis00, Maruyama04, Guerra10}.
The sticking lasts so long as the cluster-substrate surfaces are
orientationally aligned, and the long sliding jumps occur when a thermal
fluctuation rotates the cluster destroying the alignment \cite{Guerra10}.
Further understanding of the sliding of these deposited nano-objects will
be of considerable future value \cite{Schirmeisen09}.

\subsubsection{Simulated frictional control}

Exploring novel routes to achieve friction control by external physical
means is an important goal currently pursued in nanotribology.
Two methods have recently been suggested by simulation: mechanical
oscillations and phase transitions.

\begin{figure}
\centerline{
\includegraphics[width=0.48\textwidth,clip=]{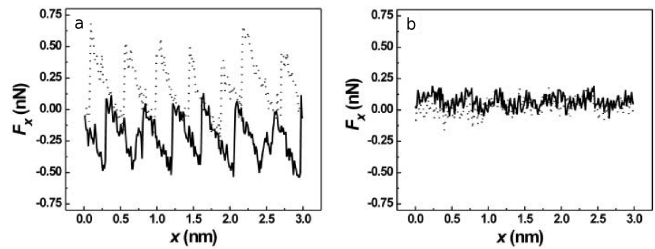}
}
\caption{\label{oscillations:fig}
The effect of oscillations on the lateral force detected by AFM scanning
forward (solid curve) and backward (dotted curve) on an atomically flat
NaCl surface.
An average normal load $F_N = 2.73$~nN was kept constant
(a) without a bias voltage between the cantilever and the sample holder and
(b) with an applied AC voltage with frequency $f=56.7$~kHz and amplitude
$1.5$~V.
From \textcite{Socoliuc06}.
}
\end{figure}

{\it Mechanical oscillations.}
Natural or artificially induced oscillations obtained by small normal or
lateral mechanical vibrations may, when applied at suitable frequency and
amplitude ranges, 
help 
driving a contacting interface out of its
potential-energy minima, increasing considerably surface mobility and
diffusion, and reducing friction and wear.
Flexibility and accessibility are the main relevant features of this
approach, since frictional properties can be tuned continuously by the
frequency and the amplitude of the applied vibrations.
This effect has been demonstrated experimentally with AFM
\cite{Socoliuc06,Lantz09,Riedo03,Jeon06}, see Fig.~\ref{oscillations:fig},
and in sheared confined system \cite{Heuberger98,Bureau00,Cochard03} and
numerically with atomistic MD \cite{Gao98,Capozza09} or MM approaches
\cite{Rozman98a,Zaloj99,Tshiprut05}.
On a larger scale, it has also been reported that in sheared granular media
experiments the stick-slip behavior is also significantly perturbed by tiny
transverse vibrations \cite{Johnson05,Johnson08,Capozza11,Krim11}.
Despite these promising numerical and experimental contributions, a
detailed analysis accounting for the friction dependence on vibrations is
still to some extent lacking.
Most past theoretical studies of mechanical control adopted an
oversimplified single-asperity model, which misses the collective behavior
of multi-asperity mesoscopic interfaces.
This and other aspects of interface oscillation are still calling for a
proper treatment and understanding.

{\it Phase transitions.}
Another idea to control friction is to employ a substrate undergoing a
phase transition.
While it is obvious that friction will change in the presence of a phase
transition, it is more subtle to qualify and quantify precisely the effect.
Surprisingly perhaps for such a basic concept, there are essentially no
experimental data available -- and no theory either.
While it would be tempting to use linear-response theory
\cite{Ala-Nissila92}, with the fluctuation-dissipation theorem making
direct contact between criticality and viscous friction, one cannot ignore
that realistic dry friction is dominated by stick-slip instabilities that
are intrinsically violent and non-linear.
Hence, one is left with MD simulations.
A PT-like MD nanofrictional simulation based on a point slider over a 2D
model substrate with a built-in structural displacive transition recently
predicted that stick-slip friction should actually peak near the substrate
critical temperature \cite{Benassi11b}.
%
These results show that friction will also depend upon the
order parameter direction, a dependence due to the different ability of the
slider to elicit disorder in the substrate depending on that direction.
%
Some level of control of atomic-scale friction
can thus be anticipated through external switching of the order parameter
direction (e.g., by an external field or strain).
Although 
the magnitude of the phase transition
effects relative to the background friction will of course depend on the real
system chosen, these results suggest pursuing this idea experimentally in,
e.g., displacive ferro and antiferrodistortive materials, a vast realm of
solids exhibiting continuous or nearly continuous structural transitions.

\section{Multicontact Models}
\label{multicontact:sec}

\begin{figure} \centering
\includegraphics[width=0.48\textwidth,clip=]{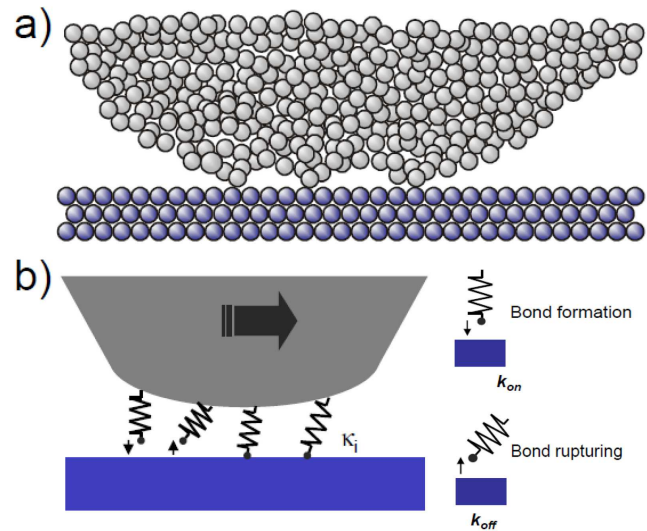}
\caption{\label{fig:mechano1}(color online).
  Multicontact modeling.
  (a) Sketch of a typical geometry for an amorphous tip sliding on a flat
  crystalline surface.
  (b) A model to simulate multiple contacts at the tip-sample interface.
  Rates of contact formation and rupturing processes are determined by the
  heights of the corresponding energy barriers, $\Delta {E_{\rm on}}$ and
  $\Delta {E_{\rm off}}$.
  Adapted from \textcite{Barel10a}.
}
\end{figure}

The PT model discussed in earlier chapters provides a good initial
description of the frictional behavior of an individual contact that can be
relevant for nanotribology experiments.
However, recent simulations \cite{Mo09} revealed that, even for an
apparently sharp AFM tip sliding on a crystalline surface, the actual
interface consists of an ensemble of individual atomic contacts (see
Fig.~\ref{fig:mechano1}).
On larger scales such as the mesoscale of SFA experiments the multicontact
picture becomes even more obvious.
The way how individual contacts can be averaged to yield macroscopic
friction law has been the focus of intense research in the past decades.
Friction is not simply the sum of single-asperity responses, but is
influenced by temporal and spatial dynamics across the entire ensemble of
contacts that form the frictional interface.
Long-range elastic interactions between contacts are important and cannot
be neglected.
\textcite{Persson01} and \textcite{Persson05} greatly improved and
generalized these concepts to more realistically fractal rough surfaces.

\subsection{Mechano-kinetic models}

A significant progress in the solution of these problems has been recently
achieved with coarse-grained mechano-kinetic models \cite{Persson95a,
  Braun02a, Filippov04, Barel10a, Barel10b, Braun09a} that describe
friction through dynamical rupture and formation of interfacial contacts
(junctions).
These contacts may represent molecular bonds, capillary bridges, asperities
between rough surfaces, and for lubricated friction they can mimic patches
of solidified lubricant or its domains.
Each contact is modeled as an elastic spring connecting the slider and the
underlying surface.
As long as a contact is intact (unbroken), it is increasingly stretched
with a speed equaling the velocity of the slider, and thus produces an
increasing force that inhibits the motion; after the instability point is
reached, a ruptured contact relaxes rapidly to its unstretched equilibrium
state.
%
%
%
The kinetics of contact formation and rupturing processes depends on
the physical nature of contacts.
For atomic scale contacts, capillary bridges and domains of solidified
lubricants, the processes of rupture and formation of contacts are
thermally activated, and the interplay between them may lead to a complex
dependence of friction on slider velocity and sample temperature
\cite{Persson95a, Braun02a, Filippov04, Barel10a, Barel10b, Braun09a,
  Braun09b}.
For micro- and macroscopic asperities between rough surfaces thermal
effects are less significant and threshold rupture forces should be taken
from a distribution that is determined by the structure of the contacting
surfaces.
The mechanism of contacts detachment is similar to the one proposed
previously by the fiber-bundle models \cite{Alava06}.

At the nanoscale the rates of formation $k_{\rm on}$ and rupturing $k_{\rm
  off}$ of microscopic contacts are defined by the corresponding energy
barriers $\Delta E_{\rm on}$ and $\Delta E_{\rm off}$.
The barrier for detachment, $\Delta E_{\rm off}$, is force dependent
and diminishes as the force acting on the contact increases and the contact
is stretched.
As we discussed above, precisely this mechanism characterizes the PT model,
but what has rarely been rationalized so far is that the process of contact
formation must be considered as well.

The dynamics of friction in the mechano-kinetic models is determined by
four characteristic rates: (i) the rate $k_{\rm off}^0$ of spontaneous
detachment of contacts, (ii) the rate $k_{\rm on}$ of contact formation,
(iii) the rate $Kv/{f_s}$ of forced unbinding, and (iv) a characteristic
rate of the pulling force relaxation, $\omega_m = \max \left( K/(\gamma m)
,\;\sqrt {K/m} \right)$.
Here $f_s$ is an average threshold force for the contact rupture, $K$ is
the stiffness of the pulling spring, $\gamma$ is the dissipation constant,
and $m$ is the mass of the slider.
Correspondingly, these models exhibit three regimes of motion:
(i) smooth sliding at very low velocities or high temperatures, with
$k_{\rm off}^0 > Kv/{f_s}$,
(ii) smooth sliding also for high velocities or low temperatures, with
$k_{\rm on} < Kv/{f_s}$, and
(iii) stick-slip oscillations for intermediate velocities and temperatures.
These multicontact models demonstrate that the overall smooth sliding
corresponds to uncorrelated atomic-scale stick-slip (or smooth)
motion of individual junctions, while the global stick-slip motion emerges
from a cooperative behavior of the junction subsystems.
It is interesting to note that the transition from smooth sliding to
stick-slip with increasing $v$ was indeed observed in SFA experiments with
two weakly adhering boundary-lubricated surfaces \cite{Drummond03}.

An important advantage of the mechano-kinetic models is that they are
directly scalable to meso and macro-scales.
Application of these models already allowed to solve some significant
disagreements between the experimental observations and the predictions of
the PT model and MD simulations.
Firstly, SFA experiments found that the critical velocity for transition
from stick-slip is in the interval of $1-10~\mu$m/s, while the MD
simulations and PT model lead to values which are 6 or 7 orders of
magnitude larger \cite{Braun02a, Luan04, Braun06}.
According to the mechano-kinetic models the transition should occur at $v
\approx f_s k_{\rm on}/K$ that for reasonable values of the parameters
agrees with the experimentally observed values of the critical velocity.
Secondly, the PT model and MD simulations \cite{Brukman08, Steiner09,
  Dong11} fail to reproduce the nonmonotonic temperature dependence of the
average friction force found in FFM experiments for several material
classes.
The mechano-kinetic model demonstrates that the peak in the temperature
dependence of friction emerges from two competing processes acting at the
interface: the thermally activated formation and rupturing of an
ensemble of atomic contacts \cite{Barel10a, Barel10b}.
This observation also provides a direct link between the temperature and
velocity dependencies of friction and it shows the experimentally observed
fingerprint in the friction-velocity curves.
Specifically, at temperatures above the peak temperature, friction
increases with scan speed, whereas, below the peak, friction decreases with
velocity.

An important and still unresolved question is what is the microscopic
origin of the rupture and reattachment processes introduced for the
interpretation of FFM experiments \cite{Filippov04, Barel10a, Barel10b}.
Similarly to the mechanism of energy dissipation in AFM
\cite{Kantorovich04, Ghasemi08}, they can be attributed to reversible jumps
of surface atoms, flips of surface fragments or transitions between
different tip structures, which are induced by the tip motion along the
surface.
These dissipative processes result in a bistable potential-energy profile
for the tip-surface junction where the barrier separating the potential
minima is continuously changed during sliding.
An unambiguous understanding of the nature of the corresponding
instabilities and evaluation of microscopic parameters which determine the
values of the rupture and reattachment rates require first-principles
calculations of potential energy surfaces for the tip-surface junctions.
The first attempts in bridging the gap between mechano-kinetic models and
MD simulations have already demonstrated that model parameters can be
completely specified using information obtained from fully atomistic
simulations \cite{Perez10}.
Then the parameter-free kinetic models are able to reproduce the
temperature and velocity variation in the friction force as obtained from
fully dynamical atomistic simulations with very high accuracy over a wide
range of conditions.
This combined approach is promising because it allows the full atomistic
details provided in MD simulations to be used in interpreting experimental
phenomena at time and length-scales relevant to tribological measurements.

\subsection{Elastic interactions and collective effects}
\label{Larkin:sec}

Elastic instabilities play, as we saw, a prominent role in explaining
frictional dissipation at the nanoscale and one may thus ask what happens
at larger scales for multicontact interfaces.
The role of elasticity in friction crucially depends on whether stress
gradient are present or not \cite{Lorenz12}.
For uniform loading, stress is distributed homogeneously on the contact
surface and elastic interactions mediate the response of local slip
fluctuations.
Experimental conditions, however, often lead to shear stress concentration
at the edge of the sample leading to detachment fronts. We discuss the
first case in this section and the second in the the next one.

The interplay between small scale disorder due to random contacts and
elastic interactions between the contacts is a complex statistical
mechanics problem that is encountered in different contexts from vortex
lines in superconductors \cite{larkin1979}, dislocations in solids
\cite{labusch1970}, domain walls in ferromagnets \cite{hilzinger1976} to
name just a few.
Two bodies in contact form a set of $n$ random contacts per unit area.
In the limit of small uniform loads, we can consider a weak-pinning
hypothesis, in which friction results from the fluctuations of the forces
due to individual contacts, and derive a scaling theory
\cite{caroli1998,volmer1997,persson1999,sokoloff2002}.
Neglecting for simplicity tensor indices, the displacement of a contact
at $\mathbf{x}$ due to the elastic interactions with the other contacts can
be estimated as
\begin{equation}\label{Green:eq}
u= \int d^2 x' G(\mathbf{x}-\mathbf{x'}) \, \sigma(\mathbf{x'})
\,,
\end{equation}
where the elastic Green function scales as $G(\mathbf{x}) \sim
1/(E|\mathbf{x}|)$, $E$ is the Young modulus (which is typically $10-100$
times larger than the yield strength $\sigma_Y$),
and the contact stresses $\sigma$ are randomly distributed with zero mean
correlations
\begin{equation}
 \langle \sigma(\mathbf{x}) \sigma(\mathbf{x'})\rangle =
 n \sigma_Y^2 a^4 \delta(\mathbf{x}-\mathbf{x'})
\,,
\label{eq:stress_corr}
\end{equation}
where $a$ is the contact diameter, depending on the normal load, the yield
stress $\sigma_Y$ is taken as a measure of the stress due to the contacts,
and $n$ is the contact density.
In analogy with other collective pinning theories, we can estimate the
typical lengthscale $L_c$ [usually referred to as the ``Larkin length''
  \cite{larkin1979} or ``elastic coherence length''
  \cite{persson1999,Muser04}] at which elastic interaction become important
by the condition that the typical displacement equals the size of the
contact $\langle u^2 \rangle = a^2$.
A straightforward calculation yields
\begin{equation}
L_c = a \exp \left(C\frac{E d}{\sigma_Y a}\right)^2,
\label{eq:larkin}
\end{equation}
where $C$ is a numerical constant, and $d =1 /\sqrt{n}$ is the average
distance between contacts.
The Larkin length separates the lengthscales for which elastic interaction
dominates ($L < L_c$) and the contact interface does not deform, from those at
which disorder dominates ($L>L_c$) and interface adapts to the pinning-center
landscape.
For two macroscopic bodies in contact, Eq.~\eqref{eq:larkin} predicts that
$L_c$ is very large because $d \gg a$ and $E \gg \sigma_Y$, and
pinning-induced deformations should be absent.
One should bear in mind that boundary-induced stress gradient can still
lead to observable elastic deformations, as discussed in the next section.

An important aspect that is missing from the analysis above is the presence
of long-range correlations: most contact surfaces are self-affine over
several lengthscales.
This implies that the assumption of uncorrelated surface stresses made in
Eq.~\eqref{eq:stress_corr} is not valid.
One can, however, repeat  the Larkin argument
for long-range correlated pinning stresses
\begin{equation}
 \langle \sigma(\mathbf{x}) \sigma(\mathbf{x'})\rangle =
\sigma_Y^2 \left(\frac{a}{|\mathbf{x}-\mathbf{x'}|}\right)^\gamma
,
\label{eq:stress_corr2}
\end{equation}
where $\gamma$ is a scaling exponent.
Computing the typical displacement in this case, we obtain
\begin{equation}
L_c \propto a\left(\frac{E}{\sigma_Y}\right)^{\frac{2}{2-\gamma}}
\label{eq:larkin2}
\end{equation}
for $\gamma <2$, while for $\gamma\geq 2$ long-range correlations are
irrelevant \cite{Fedorenko06} and we recover the uncorrelated case in
Eq.~\eqref{eq:larkin}.
The interesting feature of Eq.~\eqref{eq:larkin2} is that the dependence is
not exponential and the contact interface can deform even at small
length scales.
The effect of elastic interactions due to the contact of self-affine
surfaces  has been studied analytically and
numerically, revealing that the solid indeed deforms elastically due to the
contact \cite{persson2006,Campana2008,Campana2011}.

\begin{figure} \centering
\includegraphics[width=0.48\textwidth,clip=]{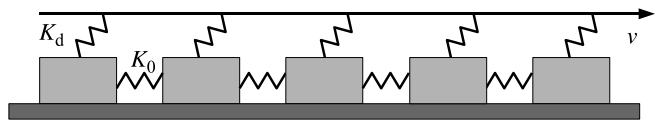}
 \caption{\label{fig:BKmodel}
 A sketch of the Burridge-Knopoff model.
 A set of frictional blocks connected by springs of stiffness $K_0$ are
 attached to a slider moving at velocity $v$ by a set of springs of
 stiffness $K_d$.
}
\end{figure}

The role of elasticity in friction is vividly illustrated by the
Burridge-Knopoff model for earthquakes \cite{burridge1967,carlson1994}
where a set of frictional blocks of mass $M$ coupled by springs are driven
over a substrate (see Fig. \ref{fig:BKmodel}).
In one dimension, the equation of motion for the displacement $u_i$ of
block $i$ is given by
\begin{equation}
M \ddot u_i = K_0(u_{i+1}+u_{i-1}-2u_i) + K_d(u_i-i a-v t) + f(\dot u_i)
\end{equation}
where $K_0$ and $K_d$ are the stiffnesses of the springs connecting the
blocks between themselves and with the loading plate that moves at constant
velocity $v$.
Here $f(v)$ is a phenomenological friction force that decreases with
the velocity of the block and $a$ is the rest length of the springs
connecting the blocks.

The Burridge-Knopoff is a deterministic model but displays a very rich
dynamical behavior with widely distributed slip events.
The key to the instability is the velocity-weakening constitutive law
$f(v)$ employed to describe the frictional properties of each block.
The model thus operates at a macroscopic scale and one needs to justify
microscopically the origin of its constitutive law.
In this context the model serves to illustrate the concept of localized
instabilities during friction: on the tectonic scale slip is localized on
some portions of the fault that does not necessarily move coherently as a
rigid block.

\subsection{Mesoscale friction: detachment fronts}
\label{detachment:sec}

Significant progress in understanding the relationship between the dynamics
of individual contacts and macroscopic frictional motion has been achieved
with the development of a new real-time visualization method of the net
area of contact along the entire interface, as discussed in a series of
papers by Fineberg's group \cite{Rubinstein04, Rubinstein06,
  Rubinstein2007, Bendavid2010, Bendavid2011}.
In their experimental apparatus, two poly(methyl methacrylate) (PMMA)
blocks were pressed together and sheared by a constant force.
A similar visualization technique has been used for tribological studies of
a different transparent material, Columbia Resin \cite{Nielsen10}.
Owing to the transparency of the media, it was possible to record the
contact area as the block were slipping.
%
This method has enabled a number of key conclusions on the mechanism of
transition from static to kinetic friction in macroscopic systems to be
drawn:
(i) the onset of sliding is preceded by a discrete sequence of crack-like
precursors (collective modes of the entire ensemble of asperities);
(ii) the transition is governed by the interplay between three types of
fronts: sub-Rayleigh, intersonic and slow fronts; and
(iii) a sequence of ``precursor'' events gives rise to a highly
inhomogeneous spatial distribution of contacts before the overall sliding
occurs.
The collective behavior of the asperity ensemble that composes a frictional
interface therefore determines the transition mechanism from static to
kinetic friction.
Imaging the contacts during friction, also allowed to record the local
values of shear and normal stresses.
By doing this, \textcite{Bendavid2010} showed that the friction coefficient
is not a constant material property but it also depends on the way the
system is loaded locally \cite{Bendavid2011}.
These results call into question many assumptions that have ruled friction
for centuries: if the onset of sliding occurs by the propagation of fronts
then elastic deformations in the contact interface become relevant and
should be taken into account by a theory of friction.
This may even call into question the Amontons law, stating that the
friction coefficient is independent on the normal load and on the apparent
contact area.
%
%
Indeed deviations from Amontons' laws have been reported for this
experiment.

The experimental phenomenology is well captured by a minimal spring-block
model \cite{Braun09a} that describes friction at the slider-substrate
interfaces in terms of rupture and reattachment of surface junctions, which
represent asperities between rough surfaces.
Contrary to the above discussed Burridge-Knopoff model of earthquakes where
phenomenological friction laws have been introduced, the model of
\textcite{Braun09a} explicitly includes two most relevant material
properties: interfacial elasticity and thresholds for yielding/rupture of
surface junctions.
The interfacial elasticity, which was ignored in previous commonly used
models of friction, defines a new velocity scale that is independent of the
Rayleigh speed and corresponds to slow crack-like fronts
\cite{Rubinstein04, Rubinstein06, Nielsen10} mediating the transition from
static to kinetic friction.

The spring-block model \cite{Braun09a} motivated a continuum description of
friction between spatially extended materials that includes a coupling
between the bulk elastic deformations and the dissipative dynamics at
frictional interfaces, which mimic the rupture and reattachment of
microcontacts \cite{Bouchbinder11,BarSinai12}.
This promising approach may bridge a gap between microscopic and
macroscopic scales and enable simulations of macroscopic friction processes
at time and length scales relevant to tribological measurements.

The 1D spring-block models discussed above
\cite{burridge1967,carlson1994,Braun09a} have limitations, which do not
allow a quantitative description of macroscopic friction experiments.
In particular, the 1D models predict an exponential decay of elastic
interactions within the sample, while a 3D description leads to a power law
decrease of the stress.
Nevertheless, recent 2D calculations employing spring-block model
\cite{Tromborg11} and finite-element method \cite{Kammer12} demonstrated
that the 1D models provide an important insight into the mechanism of dry
friction between spatially extended materials, and allow to investigate the
effect of system parameters on frictional response.
It is important to note that, in order to reproduce experimentally observed
spectrum of detachment fronts, the models should incorporate interfacial
stiffness and local friction laws including both velocity weakening and
strengthening branches \cite{Braun09a,Bouchbinder11,BarSinai12}.
The models assuming local Coulomb-Amontons friction at the block-substrate
interface \cite{Tromborg11} do not exhibit slow rupture fronts like those
observed in various experiments \cite{Rubinstein04,
  Rubinstein06,Nielsen10}.
A quantitative comparison to experimental data requires 2D calculations
with a proper choice of local friction laws that is a challenge for future
studies.

Several questions still remain open: how general are these experimental
results?
Do they depend on the specific material, or the setup geometry?
If not, should we revise our general understanding of friction based on
Amontons' laws?
Yet Amontons' laws have worked quite well for centuries so they should still be
valid at least in an average sense or in some conditions.
All these questions require further theoretical insight and more experiments
in mesoscale friction.

\section{A few special frictional phenomena}
\label{special:sec}

\subsection{Electronic friction}

As many theorists \cite{Schaich81, Persson96, Novotny99, Liebsch99,
  Plihal99, Sokoloff00, Persson00a, Sokoloff02,Volokitin07} have discussed,
sliding friction over a metallic substrate should elicit electronic
excitations, giving rise to additional frictional dissipation besides that
due to phonons.
Thus for example friction on a metal surface should drop when the metal is
cooled below the superconducting critical temperature, as normal
electron-hole gapless excitations disappear.
A first QCM report of this frictional drop for the sliding of molecular
N$_2$ islands on a Pb surface \cite{Dayo98} broke the ice, but also
triggered a considerable debate \cite{Renner99,Krim99,Fois07}.
More recently, electronic friction was demonstrated in doped
semiconductors, where local carrier concentration was controlled through
application of forward or reverse bias voltages between the AFM tip and the
sample in the {\it p} and the {\it n} regions, thus demonstrating the
capability to electronically tune friction in semiconductor devices
\cite{Park06}.

\begin{figure} \centering
\includegraphics[width=0.47\textwidth,clip=]{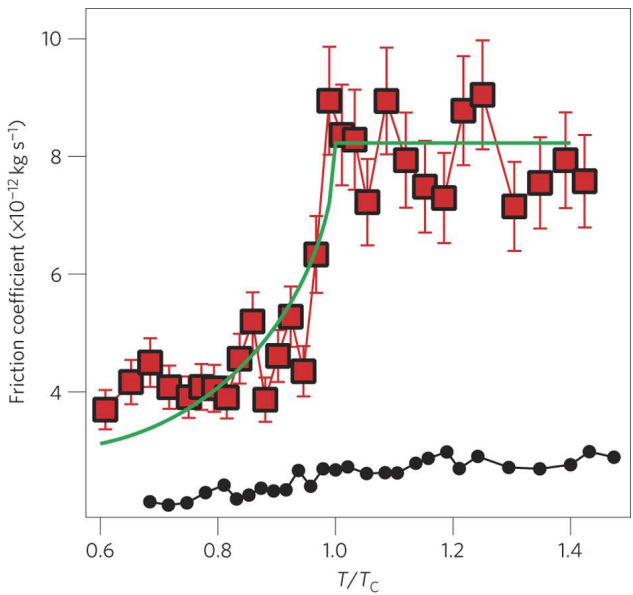}
 \caption{\label{fig:Kisiel11_nature_mat}
 Temperature variation of the friction coefficient across the
 critical point $T_c=9.2$~K of Nb.
 Red squares / black dots: tip-sample separation 0.5~nm / several $\mu$m
 (free cantilever).
 Green solid line: fit by the analytic dependency expected from the
 Bardeen-Cooper-Schrieffer theory.
 The friction coefficient is shifted vertically by $2.5\times
 10^{-12}$~kg~s$^{-1}$.
 From \textcite{Kisiel11}.
}
\end{figure}

Very recently, using a pendulum-type AFM probe, a clear noncontact friction
drop over the surface of Nb has been characterized at the superconducting
transition \cite{Kisiel11}.
The features observed at this transition follow quite closely the
predictions by Persson and collaborators,
see Fig.~\ref{fig:Kisiel11_nature_mat}.
This ultra-sensitive pendulum probe is now ready to be put to work to
detect the change of electronic friction at other superconducting
transitions.
The main progress to be expected is now experimental, more than
theoretical.
Promising cases should include the high-$T_c$ cuprates \cite{Damascelli03},
organics \cite{Kanoda06}, fullerides \cite{GunnarssonBook,Capone09},
heavy-fermion compounds \cite{Ernst10} and pnictides \cite{Yin09}.
One interesting physical issue that these investigations could address is
the nature and role of strong electron correlations, generally believed to
be important for superconductivity in most of these materials.
Strongly-correlated superconductivity models generally imply that, when not
superconducting, the metal is not always Fermi-liquid like, therefore with
vanishingly small quasiparticle strengths and Drude weights.
Moreover, it should be noted that a recent experimental and theoretical
study demonstrated the general existence at a strongly correlated metal
surface of an insulating ``dead layer'' with a thickness which for example
in V$_2$O$_3$ reaches several nanometers \cite{Rodolakis09,Borghi09}.
Another interesting case to be studied could be two dimensional
metal-insulator transitions at some surfaces, such as the Mott transition
reported for Sn/Si(111) $\sqrt 3\times\sqrt 3$ \cite{Modesti07,Profeta07}.

\subsection{Magnetic dissipation}

The relationship between nanofriction and magnetism at the atomic level is
an intriguing side direction.
In a recent Magnetic Exchange Force Microscopy (MExFM) experiment
\cite{Wiesendanger09}, atomic force sensitivity to the magnetic state of a
surface atom was demonstrated for an atomically sharp Fe magnetic tip over
the $(001)$ surface of antiferromagnetic NiO \cite{Kaiser07}.
Besides a different force for the two oppositely polarized surface Ni atoms
-- well explained by the Fe-Ni exchange available from electronic structure
calculations \cite{Momida05} -- the results show a surprising difference of
mechanical dissipation, with a 
large
excess of order 15-20~meV per
cycle in the antiparallel Fe-Ni spin configuration, as compared to the
parallel one.
The Fe-Ni exchange energy is higher in the antiparallel case, and the
difference can clearly be dissipated by flipping the Ni spin.
However,
direct excitation of surface antiferromagnetic magnons in the antiparallel
tip-Ni case -- the first obvious possible explanation -- is ruled out
since, owing to strong dipolar anisotropy, the antiferromagnetic spin-wave
dispersion of NiO has a gap $\Delta\sim1.5$~meV $\sim 0.36$~THz
\cite{Hutchings72} in bulk, and one at least as large at the surface.
The low-frequency oscillatory perturbation exerted by the tip ($\sim
160$~KHz) on the surface spin, far smaller than this gap, is completely
adiabatic, and direct dissipation in the spin-wave channel is impossible.
Other strictly magnetic dissipation mechanisms involving mesoscopic scale
phenomena, such as domain wall motion \cite{Liu97}, also appear inefficient
in the atomic scale tip-sample magnetic dissipation.
One is left with magnetic coupling to surface atomic displacements.
Acoustic phonons are not gapped, both in bulk and at the surface, so they
could indeed dissipate.
However, the probe frequency is extremely low; and since dissipation
vanishes in linear response theory as a high power of frequency
\cite{PerssonTosatti99}, a magnetic dissipation mechanism via phonons must
involve some hysteretic phenomena far from linear response.
Recent theory work \cite{Pellegrini10} suggests that the nonlinear response
may be related to the attainment of a strong coupling overdamped
spin-phonon state very well known in other contexts \cite{Caldeira81},
giving rise to an unusual kind of single-spin hysteresis.
The tip-surface exchange interaction couples together spin and atom
coordinates, leading to a spin-phonon problem with Caldeira-Leggett type
dissipation.
In the overdamped regime, that coupled problem can lead to a unique
single-spin hysteretic behavior with a large spin-dependent dissipation,
even down to the very low experimental tip oscillation frequencies, just as
is seen experimentally.
A quantum phase transition to an underdamped regime with a loss of
hysteresis and a dramatic drop of magnetic tip dissipation should in
principle be found by increasing and tuning the tip-surface distance.
This experimental check would also help distinguish this interesting
spin-phonon mechanism from more trivial possibilities, such as a additional
dissipation simply due to a closer distance Fe tip -- NiO surface approach
in the antiparallel spin configuration.

\subsection{Carbon nanotube friction}

Carbon and carbon nanotubes (CNT) are widely employed in tribology
\cite{Chen03}.
Nanotube applications are also numerous in nanofriction.
\textcite{Falvo99} and also \textcite{Buldum99} discussed the possibility
to slide or roll nanotubes on a surface.
Recently, a large longitudinal-transverse anisotropy of AFM friction on
surface deposited CNTs has been explained precisely by the contrast
between the longitudinal tube rigidity, against the transverse softness
manifested 
by ``hindered rolling'' \cite{Lucas09}.
%

Zettl and collaborators \cite{Cumings00,Kis06} demonstrated ultralow
friction experienced by coaxially sliding multiwall CNTs.
Coaxial CNT sliding also inspired numerous simulations
\cite{Servantie03,Servantie06a,Servantie06b}.
More simulation identified curious -- even if at present rather academic --
frictional peaks at selected sliding speeds \cite{Tangney06} corresponding
to the parametric excitation of quantized nanotube breathing modes.
Recent theory work \cite{Zhang09} also discovered that the twofold
degeneracy of the breathing modes can cause chiral symmetry to break
dynamically at these frictional peaks, so that even purely longitudinal
coaxial sliding of nonchiral tubes can generate angular momentum.
Other mechanical/rheological properties of CNTs have also been
probed by AFM \cite{Palaci05}.

\begin{figure}
\centerline{
\includegraphics[width=0.48\textwidth,clip=]{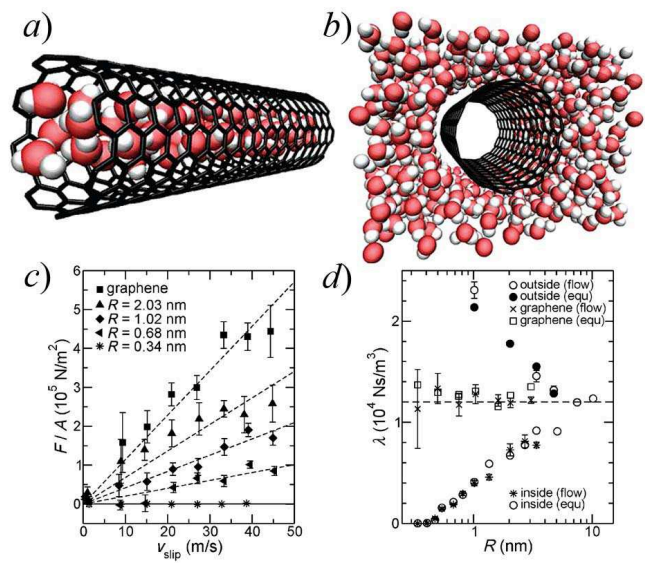}
}
\caption{\label{CNTwater}(color online).
Simulated flow of water (a) inside or (b) outside armchair CNTs, and the
ensuing friction dependency on (c) slip velocity and (d) confinement radius
$R$.
$F_k$ is the kinetic friction force, and $\lambda$ is the friction
coefficient.
Panels (c) and (d) include data for graphene slabs, for which $2R$ is the
wall-to-wall distance.
Adapted from \textcite{Falk10}.
}
\end{figure}

Water wetting and the interfacial friction of water in CNTs have also
been studied for various purposes.
The observation of flow-generated voltages growing logarithmically with
velocity of ion-rich water \cite{Ghosh03} appears to be a manifestation of
electronic friction which has found various explanations including
stick-slip of ions embedded in high viscosity water near the tube
\cite{Persson04b} or a statistical consequence of the flow-induced
asymmetry in the correlation of the ions, in the ambient fluid as seen by
the charge carriers in the CNT \cite{Ghosh04}.

Recent studies have focused on disentangling confinement and curvature
effects on water friction inside and outside CNTs, showing that the
friction coefficient exhibits a strong curvature dependence.
While for a flat graphene slab friction is independent of confinement, it
decreases with CNT radius for water inside, but increases for water
outside, see Fig.~\ref{CNTwater}.
The friction coefficient is found to vanish below a threshold diameter for
armchair CNTs.
A ``superlubric'', structural origin of this curvature dependence
associated with a curvature-induced incommensurability between the water
and carbon structures, has been proposed \cite{Falk10}.

\subsection{Friction in colloidal systems}

Handling matter with static periodic fields generated by interfering lasers
adds a new exciting actor to the realm of toy systems that display real
physics.
Trapping and handling colloidal particles with intense photon fields offers
the possibility to change parameters freely, to compare directly experiment
with theory, to test theoretical predictions, and to visualize directly in
simple cases the intimate mechanics of sliding friction
\cite{Vanossi12,Vanossi12PNAS}.
In the novel approach freshly inaugurated by Bechinger's group
\cite{Bohlein12}, a 2D close-packed crystal of charged colloidal particles
is forced to slide by Stokes forces in the presence of a laser-generated
static potential.
Unlike conventional sliding crystal surfaces, two-dimensional lattices with
different symmetries, lattice spacings and corrugation amplitudes can be
constructed at will, realizing for example commensurate or incommensurate
matchings, quasicrystal substrate geometries, and possibly ``disordered''
geometries too.

Besides and above all that, colloidal friction provides an unprecedent
real-time visualization of the full frictional dynamics at play.
Unlike AFM, SFA ad QCM which provide crucial, but averaged frictional data
such as the overall static and kinetic friction, mean velocities, slip
times, etc., the colloidal experiments photograph the actual time-dependent
motion of every individual particle during sliding, an exquisite privilege
restricted so far to the ideal world of MD simulations.
Transitions between different dynamical states become experimentally
accessible and can be analyzed and related to the detailed particle motion.
These and other goodies lie in the future.

\section{Conclusions}
\label{conclusions:sec}

Among provisional conclusions of this Colloquium we mention:

(\textit{i}) Despite limitations, all levels of modeling and simulation are highly informative and predictive.

(\textit{ii})
The PT model is, despite its deceiving simplicity, extremely useful in
understanding many aspects of nanofriction. Its extreme success tends
however to hide the actual complexity of the phenomenon.

(\textit{iii})
Multi-contact models are instrumental in describing mesoscopic friction and
fracture.
One main problem of these models is the multiplicity of empirical
parameters they involve.
Other open problem are under what conditions these models yield realistic
stick-slip friction as opposed to smooth sliding, and the exact role of the
elastic interactions between contacts.

(\textit{iv})
Molecular-dynamics simulations are good and informative for qualitative
descriptions of atomic stick-slip.
Open problems are the high speed of observed change from stick-slip to
smooth sliding; the potential artifacts introduced by unrealistic
dissipation mechanisms of Joule heat, and more importantly the simulation
size and time limitations, in particular the complete omission of slow,
logarithmic relaxations and ageing.

(\textit{v})
Prospective mechanisms for the control of friction, such as mechanical
oscillations or a phase transition in the substrate, are suggested by
model studies, and are presently under experimental scrutiny.

It is worth mentioning in closing that there remain fully open problems at
the very basic theoretical level.
First of all, we do not have a proper theory of friction, namely a theory
where the frictional work could be calculated quantitatively (not just
simulated) in all cases -- they are the majority -- where linear-response
theory is inapplicable.
Second, while MD simulations can to some extent be used in lieu of theory,
they have been so far strictly classical.
Future work should include quantum effects and gauge their importance.

There are many more outstanding challenges left in nanofriction. Among
them:

\begin{compactitem}

\item
Bridging the gap between nano and meso (macro) scale friction: multicontact
systems.

\item
Mechanical control of friction in multicontact systems.

\item
The ageing of surface contacts at the nano and macroscales.

\item
Role of wear and adhesion at the nanoscale.

\item
Rolling nanofriction: besides the known case of nanotubes:
does it exist, and how to distinguish between rolling and sliding?

\item
Friction in dislocations and in granular systems.

\item
Friction in biological systems (motor proteins, cells membranes and pores,
etc.).

\end{compactitem}
Lively progress along these and newer lines is to be expected in the near
future.

\acknowledgments

We wish to express our gratitude to
I.\ Barel,
C.\ Bechinger,
A.\ Benassi, A.\ R.\ Bishop,
T.\ Bohlein,
O.\ M.\ Braun, R.\ Capozza,
D.\ Ceresoli, F.\ Ercolessi,
A.\ E.\ Filippov, J.\ Fineberg,
R.\ Guerra,
H.\ Haeffner,
J.\ Klafter,
E.\ Meyer,
G.\ Paolicelli, F.\ Pellegrini,
B.\ N.\ J.\ Persson,
T.\ Pruttivarasin,
E.\ Riedo, G.\ E.\ Santoro, A.\ Schirmeisen, U.\ Tartaglino, N. Varini,
X.\ H.\ Zhang, and T.\ Zykova-Timan
for collaboration and helpful discussions.
This work is partly funded by
the Italian Research Council (CNR) and Israel Science
Foundation via Eurocores FANAS,
the Italian Ministry of University and Research through PRIN
No. 20087NX9Y7 and No. 2008Y2P573,
the Swiss National Science Foundation Sinergia CRSII2\_136287,
and by the ComplexityNet pilot project LOCAT.

%

\end{document}